\documentclass[10pt,letterpaper]{article}
\usepackage[top=0.85in,left=1.0in,footskip=0.75in,marginparwidth=1in]{geometry}

\usepackage[right]{lineno}

\usepackage{microtype}
\DisableLigatures[f]{encoding = *, family = * }

\usepackage{enumitem}
\usepackage{multicol}
\usepackage{multirow}

\usepackage{listings}
\usepackage{amsmath}
\usepackage{amssymb}
\usepackage{IEEEtrantools}
\usepackage{booktabs}

\usepackage{xparse}

\let\doleft\left
\let\doright\right
\NewDocumentCommand{\E}{s d[]}{%
    \operatorname{E}%
    \IfValueT{#2}{\IfBooleanF{#1}{\doleft[}%
    #2%
    \IfBooleanF{#1}{\doright]}}
}
\NewDocumentCommand{\Var}{s d()}{%
    \operatorname{Var}%
    \IfValueT{#2}{\IfBooleanF{#1}{\doleft(}%
    #2%
    \IfBooleanF{#1}{\doright)}}
}
\NewDocumentCommand{\PR}{s r()}{%
    \operatorname{P}\doleft(%
    \IfValueTF{#1}{\text{#2}}{#2}%
    \doright)%
}

\raggedright
\usepackage{changepage}

\newcommand{\figfmt}[1]{\textsc{Figure #1}}
\newcommand{\tabfmt}[1]{\textsc{Table #1}}
\newcommand{\eqfmt}[1]{\textsc{Equation (#1)}}

\usepackage[aboveskip=1pt,labelfont=bf,labelsep=period,singlelinecheck=off]{caption}

\usepackage[round]{natbib}
\usepackage{nameref}
\usepackage[hidelinks]{hyperref}

\usepackage{lastpage,fancyhdr,graphicx}
\fancyheadoffset[L]{2.25in}
\fancyfootoffset[L]{2.25in}

\usepackage{color}

\definecolor{Gray}{gray}{.25}

\usepackage{graphicx}

\usepackage{sidecap}

\usepackage{wrapfig}
\usepackage[pscoord]{eso-pic}
\usepackage[fulladjust]{marginnote}
\reversemarginpar

\usepackage{textcomp}

\begin{document}
\vspace*{0.35in}

\begin{flushleft}
{\Large
\textbf\newline{A rodent paradigm for studying perceptual decisions under asymmetric reward}
}
\newline
\\
Xiaoyue Zhu \textsuperscript{1,2,4},
Jeffrey C. Erlich\textsuperscript{1,2,3,4,5*}
\\
\bigskip
\bf{1} NYU-ECNU Institute of Brain and Cognitive Science at NYU Shanghai, China
\\
\bf{2} NYU Shanghai, Shanghai, China
\\
\bf{3} Shanghai Key Laboratory of Brain Functional Genomics (Ministry of Education), East China Normal University, Shanghai, China 
\\
\bf{4} Center for Neural Science, New York University, New York, NY, United States of America
\\
\bf{5} Sainsbury Wellcome Centre for Neural Circuits and Behaviour, University College London, London, United Kingdom
\\

\bigskip
* jerlich@nyu.edu

\end{flushleft}

\section*{Abstract}
Many real-life decisions involve both perceptual processes and weighing the consequences of different actions. However, the neural mechanisms underlying perceptual decisions have typically been examined separately from those underlying economic decisions. Here, we trained rats to make choices informed by both perceptual and value cues on a trial-by-trial basis. As in typical perceptual tasks, subjects were rewarded for correctly categorizing a tone relative to a learned threshold. To add an economic component, a light indicated, on each trial, whether correct responses to one side gave higher rewards than correct responses to the other side. As such, on trials with some perceptual uncertainty, it could be worthwhile to choose the unlikely option, if it had higher expected value.  We found that, despite subjects sensitivity to the frequency of the cue and the reward sizes, their behavior was not optimal: subjects tended to shift their choices in a stimulus-independent way following light flashes. Moreover, subjects tended to under-shift, which could be interpreted as being over-confident in their perceptual beliefs or as being risk-averse. 





\nolinenumbers

\newpage

\section*{Introduction}

We are often required to make decisions based on noisy perceptual evidence where the costs associated with one choice are quite different than the costs associated with the other choice. 
Consider a dramatic example: a radiologist who needs to detect the presence (or absence) of a malignant tumor from a CT scan. The judgment itself should only be based on perceptual information, such as the density, shape and curvature of the anomalous cluster. However, the cost of mistakes in these medical decisions are highly asymmetric. If the tumor is present and the doctor says `no' (miss), the patient will be discharged and possibly die from the lack of treatment. 
On the other hand, if the tumor is absent and the doctor says `yes' (false alarm), the patient will just go through more tests until the determination of cancer is clear. When a perceptual judgment incurs asymmetric outcomes, the decision maker must integrate both the strength of sensory evidence \textit{and} the distinct costs of mistakes, or different benefits of actions. In fact, such decisions are commonplace in the real-life, from judging if the yogurt has gone bad to an animal judging whether the noise is caused by a predator or a prey. 

Laboratory tasks studying the cognitive and neural mechanisms of decision under noisy sensory information typically assign equal reward to the options, where the subject is rewarded for categorizing the stimulus correctly. 
Recently, there is an emerging interest in studying perceptual decisions with asymmetric costs and reward, predominantly in humans \citep{diederichModelingEffectsPayoff2006, diederichFurtherTestSequentialsampling2008, summerfieldEconomicValueBiases2010, gaoDynamicIntegrationReward2011, mulderBiasBrainDiffusion2012} and non-human primates \citep{fengCanMonkeysChoose2009, rorieIntegrationSensoryReward2010}. 
Despite the differences in tasks and models used, all of these experiments investigated how value information was integrated with sensory information during the decision process. One plausible mechanism is that the value information affects the processing of sensory information, such as by directly modulating the activity in primary sensory areas. \citet{stanisorUnifiedSelectionSignal2013} found reward value is a good predictor of monkey V1 activity in a curve-tracing task, likely mediated by the top-down control of attention. Similar evidence was found in human V1, whose activity was modulated by reward value even in the absence of an overt saccade \citep{serencesValueBasedModulationsHuman2008}. 

On the other hand, value information can also influence perceptual choices by adjusting the starting or ending point of the decision process. Evidence for this view came from studies using variants of the drift diffusion model (DDM), which depicts the decision mechanism as a `diffusion' process, where the decision variable `drifts' towards a threshold based on upcoming sensory information \citep{ratcliffTheoryMemoryRetrieval1978}. Naturally, the model's way to reflect an asymmetric starting point (as a result of value) would be to change the starting position for the decision variable. Such model-based analysis has shown that a shift in the starting position of DDM can best explain behavior in human subjects \citep{summerfieldEconomicValueBiases2010, gaoDynamicIntegrationReward2011, mulderBiasBrainDiffusion2012} and non-human primates \citep{rorieIntegrationSensoryReward2010}.

These two alternative hypotheses, that value information exerts influence \textit{on} or \textit{separate\ from} sensory processing, predict that asymmetric reward should lead to differential neural activity in the sensory or secondary motor areas, respectively. Moreover, causal evidence for either hypothesis can be obtained by inactivating the candidate areas during the stimulus presentation or choice phase. Th rat is an excellent model organism for studying the neurobiology of decision-making. Not only it is cost-effective, it also allows for manipulations with high temporal and spatial precision that are otherwise difficult in primates \citep[e.g.][]{deisserothCircuitDynamicsAdaptive2014, kramerOptogeneticPharmacologyControl2013}.
Numerous groups have demonstrated that rats can learn complex perceptual and economic decision-making tasks, guided by visual and auditory cues \citep{constantinopleAnalysisDecisionRisk2019, millerDorsalHippocampusContributes2017, erlichDistinctEffectsPrefrontal2015, zhuFrontalNotParietal2021}. \citet{lakDopaminergicPrefrontalBasis2020} trained mice on a task where they detected visual gratings with varying contrast, shown on the left or right monitor. The reward was asymmetric such that in alternating blocks, reporting one side correctly entailed a larger reward than the other side. To maximize reward, the animals must integrate reward history with trial-by-trial visual cues. This is the first rodent task, as far as we are aware of, that investigates percept-value integration in a decision-making context. However, as the reward structure was not explicitly cued on each trial, this task is better suited for studying the learning of action-values than percept-value integration \citep{behrensLearningValueInformation2007}. It is difficult to know exactly when in the trial the integration may be happening.
Thus, we set out to develop a rodent task where the subject's choice is guided by both the perceptual and value-based components on a trial-by-trial basis. 

\section*{Materials and Methods}

\subsection*{Subjects}
Data from 7 male rats (4 Brown Norway, 3 Sprague Dawley; Vital River, Beijing, China) is included in this study. The animals were placed on a controlled-water schedule and had access to free water 20 minutes each day in addition to the water they earned in the task. They were kept on a reversed 12 hour light–dark cycle and were trained during their dark cycle. Animal use procedures were approved by New York University Shanghai International Animal Care and Use Committee following both U.S. and Chinese regulations.

\subsection*{Behavioral Apparatus}
Animal training took place in custom behavioral chambers, located inside sound- and light-attenuated boxes. Each chamber (23 x 23 x 23 cm) was fitted with 8 nose ports arranged in four rows (\figfmt{\ref{fig:behavior}A}), with a pair of speakers on the left and right side. Each nose port contained a pair of blue and yellow light emitting diodes (LED) for delivering visual stimuli, as well as an infrared LED and infrared phototransistor for detecting rats' interactions with the port. The port in the bottom row contained a stainless steel tube for delivering water reward. Animals were loaded and unloaded from the behavioral chambers by technicians daily on a fixed schedule. Each training session lasted for 90 minutes. 

\subsection*{The perceptual gambling task}
Trials began with both yellow and blue light-emitting diodes (LED) turning on in the center port. This cued the animal to poke its nose into the center port and hold it there for 1 s -- the `fixation' period. As soon as the animal started fixation, a 500 ms tone would play from both speakers. The tone's frequency (in $\log_2(kHz)$ space) was sampled from a Gaussian distribution centered at 3 and truncated at 2 and 4, values corresponding to 8 kHz, 4 kHz and 16 kHz. Unless otherwise specified, we will use the $\log_2(kHz)$ value throughout this manuscript. Specifically, the probability density function, $\psi$, describing the distribution of $\log_2(kHz)$ tone frequencies was:   

\begin{align}
    \psi (\mu, \sigma_s, \alpha, \beta; s) &=
    \begin{cases}
        0,\ \mathrm{if} \ s < \alpha \\
        \mathcal{N} \sim (\mu, \sigma_s),\ \mathrm{if} \ \alpha \leq s \leq \beta \\
        0,\ \mathrm{if} \ s > \beta \\
    \end{cases}    
\end{align}

\noindent where $\sigma_s$ is the standard deviation and controls the difficulty of the perceptual task; it was tuned for each animal. The boundaries where $\psi$ is truncated are defined by $\alpha$ and $\beta$. The perceptual task required subjects to report whether the $\log_2(kHz)$ of the tone was greater or less than 3. We counterbalanced the left / right assignment across animals, that animals with even subject ID were rewarded for tones $\le 3$ on the left, and animals with odd ID were rewarded for tones $\le 3$ on the right.
We refer the correct port for frequencies lower than 3 as the `low port', and the correct port for frequencies higher than 3 as the `high port'. After 1 s fixation, the animal was free to withdraw from the center port and poke into the left or right choice port. The animal was rewarded with the base amount if it chose correctly, no reward was delivered otherwise. If a trial had no flash and the animal was rewarded the base amount for choosing the correct port, we refer to these trials as `perceptual trials'. Around 30\% to 65\% of the total trials in a session were perceptual trials, the proportion was different for each animal ($51.4\ [30.7, 64.2]$, mean and 95\% C.I.).

On some trials, concurrent with the tone, the three ports of one side would flash their yellow LEDs in the rate of 10 Hz, lasting for the entire duration of fixation. 
The selection of the flashing side was independent of the correct side indicated by tone frequency. If the flashing side coincided with the correct side, the animal would be rewarded with $\kappa$
times of the $base~reward$ if it chose the correct port. The reward multiplier $\kappa$ was tuned for each animal. If the flashing side was different than the correct side, the animal was rewarded the base amount if it chose the correct port. No reward was delivered for choosing the incorrect port. We refer to these trials as `perceptual gambling (PG) trials'.  
Around 35\% to 70\% of the total trials in a session were PG trials ($48.5\ [35.7, 69.2]$). The inter-trial intervals (ITI) were between 3 and 10 seconds. A trial was considered a violation if the animal failed to poke into central 300 s after trial start, or it did not make a choice 30 s after fixation. Violations were excluded from all analyses. 

\subsection*{Training pipeline} 
Animal training took place in four distinct phases: the operant conditioning phase, the fixation phase, the perceptual phase and the perceptual gambling phase.

\paragraph{The operant conditioning phase} 
In the operant conditioning phase, naive rats became familiar with the training apparatus and learned to poke into the reward port when illuminated. Trials began with the illumination of reward port, and water reward was immediately delivered upon port entry. After the rats learned to poke in the reward port reliably, they proceeded to the next training stage where they had to first poke into an illuminated choice port (left or right, randomly interleaved) before the reward port was illuminated for reward. They graduated to the next phase if they correctly performed these trials at least 40\% of the session. 

\paragraph{The fixation phase} 
In the fixation phase, rats started by initiating the trial by poking into the center port. To facilitate initial learning, only two tones were presented (2 and 4) and the same tone was presented in blocks of 5 to 20 trials. The fixation duration started from 0~ms, and was increased by 5~ms every time the rat maintained fixation in the previous trial, otherwise it remained unchanged. Rats graduated to the next phase once the fixation time reached 1~s and they could reliably choose the correct port given the frequency (75\% correct rate overall).  

\paragraph{The perceptual phase} 
The goal of the perceptual phase was to train the animals on the complete range of tone frequencies. Rats started with only 2 frequencies per side ([2, 2.25] and [3.75, 4]) in blocks of 5 to 20 trials. Once they showed sensitivity ($p < 0.05$) to the existing tones based on a generalized linear model (GLM), more intermediate frequencies were added in pairs.
The complete list of discrete frequencies was [2, 2.25, 2.5, 2.75, 3.25, 3.5, 3.75, 4], as they were spaced evenly apart in $\log_2(kHz)$ space. 
Once the performance was stable on all discrete frequencies, we introduced continuous frequencies by sampling from one of two truncated Gaussian distributions: $\psi (3, \sigma_s, 2, 3; s)$ and $\psi (3, \sigma_s, 3, 4; s)$, while the correct port was the same in each block.
Initially, $\sigma_s$ was set to be large to expose the animals to a wide range of frequencies and made the task relatively easy. Once the animals displayed sharp psychometric curves with continuous stimuli in blocks, we removed the block structure and sampled from the full truncated Gaussian $\psi (3, \sigma_s, 2, 4; s)$. The rats graduated from the perceptual phase if they showed sensitivity to the tones according to the GLM.  


\paragraph{The perceptual gambling phase} 
Rats entered the final perceptual gambling phase with good understanding of the frequency-to-choice mapping. The goal of this phase was to let animals learn the meaning of the light flashes, which was introduced with a block structure. In a block of 20 to 30 trials, only one side would flash while the tone was still drawn from the truncated Gaussian distribution. In this phase, various task parameters were adapted to each animal's reward sensitivity to induce the `perceptual gambling effect'. For example, if an animal did not shift its choice to the flashing side, we would increase $\kappa$ to increase the expected value of the flashing side, decrease $\sigma_s$ to make the trials more perceptually challenging and thus increase perceptual uncertainty, or increase the block length. The block length was gradually reduced to 1 - 3 once the subjects reliably shifted their choices in response to the flashing side.

\subsection*{Modeling with Bayesian decision theory}
The perceptual gambling task is a binary classification task with asymmetric action costs. On each trial, the rat has to take an action that requires inferring the correct class $C$ of its auditory observation $x$ from the actual stimulus frequency $s$. The probability of occurrence of each class is captured by the probability distribution $p(C)$, known as class priors. The distribution of the observation is specified conditioned on the class $C$ and denoted by $p(x|C)$, this is known as the likelihood. In our task, $p(x|C)$ cannot be directly known but had to be derived from other conditional distributions, which will be described below. Together, the distributions $p(C)$ and $p(x|C)$ define a `generative model', a Bayesian description of how the observations arise from the auditory stimulus presented on each trial \citep{maBayesianDecisionModels2019}. The main assumption of Bayesian modeling is that the rat has learned the distributions specified in the generative model, and it utilizes this knowledge fully when inferring possible states of the world. This is done by using Bayes’ rule,

\begin{align}
    p(C|x) = \frac{p(x|C)p(C)}{p(x)}
\end{align}

\noindent where $p(C|x)$ denotes the inferred posterior probability of a certain class given the stimulus frequency, and $p(x)$ acts as a normalization factor. We describe our modeling process in three distinct steps: defining the generative model, computing the posterior distribution, and choosing an action to minimize cost \citep[following][]{maBayesianDecisionModels2019}. 

\begin{figure}[htbp]
\includegraphics[width=.7\linewidth]{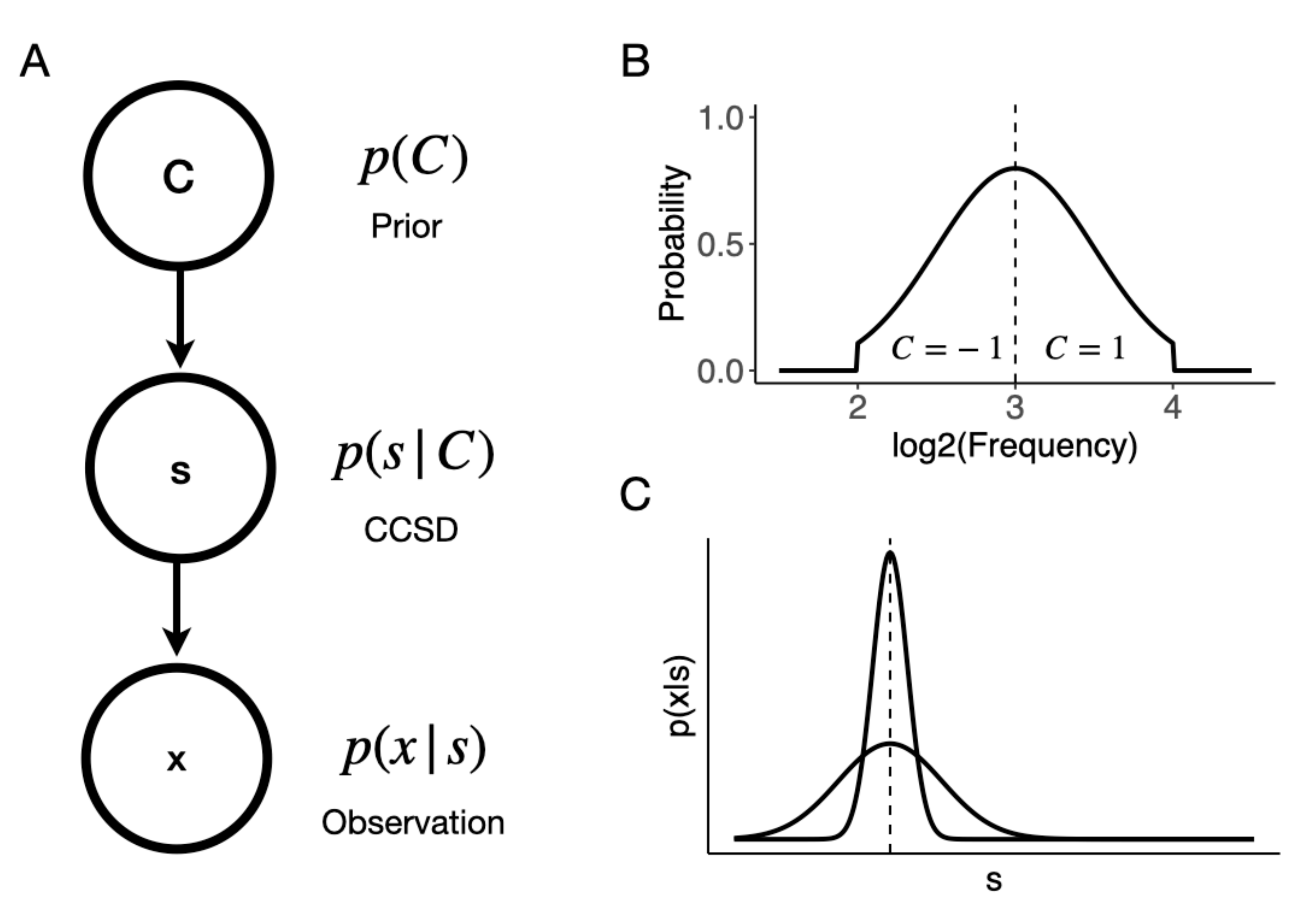}
\caption{The generative model of Bayesian inference.
\textbf{A.} Diagram of the generative model. The stimulus $s$ is drawn from the class $C$, and the observation $x$ is based on $s$ only. 
\textbf{B.} Schematic of the mirror-imaged class-conditioned class distribution in our task.
\textbf{C.} Schematic of the observation distribution $p(x|s)$ given stimulus $s$. A higher perceptual noise (larger $\sigma_p$) produced a wider distribution, whereas a lower perceptual noise (smaller $\sigma_p$) gave a more concentrated distribution around $s$.
}
\label{fig:method}
\end{figure}

\paragraph{Defining the generative model} 
The generative model is shown in \figfmt{\ref{fig:method}A}. 
It has three nodes: the correct class $C$, the stimulus frequency $s$, and the animal's noisy observation $x$. 
The rat's goal was to correctly report $C$, making that the world state of interest. 
$C$ can take on two possible values: 1 for the high port being correct, and -1 for the low port being correct. 
Associated with $C$ is a distribution $p(C)$, which is specified by two values, $p(C=1)$ and $p(C=-1)$. In our task, $p(C=1) = p(C=-1) = 0.5$, as the correct class was determined by the frequency relative to 3 drawn from the symmetrical Gaussian distribution. 
 The stimulus distribution is thus a class-conditioned stimulus distribution (CCSD), which is denoted by $p(s|C=-1)$ and $p(s|C=1)$ for the two classes, respectively. 
 As the stimulus $s$, was drawn from the Gaussian $\psi(3, \sigma_s, 2,4; s)$ and then designated to be $C=sign(s - 3)$, the two CCSDs are mirror-images of each other (\figfmt{\ref{fig:method}B}). Formally, 

\begin{align}
    p(s|C = -1) & = 
    \begin{cases}
        \psi (3, \sigma_s, 2, 3; s), \ \mathrm{if} \ 2 \leq s < 3 \\
        0,\ \mathrm{otherwise} \\
    \end{cases} \\
    p(s|C = 1) & = 
    \begin{cases}
        \psi (3, \sigma_s, 3, 4; s), \ \mathrm{if} \ 3 \leq s \leq 4 \\
        0,\ \mathrm{otherwise}
    \end{cases}
\end{align}

To complete the generative model, we need to define how the animals made observations based on the actual stimulus, $s$. Conventionally, the observation, $x$, is defined as a normal distribution centered at $s$ with standard deviation $\sigma_p$, which denotes the perceptual noise of the animal.

\begin{align}
    p(x|s) \sim \mathcal{N}(s, \sigma_p)
\end{align}

\paragraph{Computing the posterior distribution}
Computing the posterior distribution $p(C|x)$ involves both the class likelihood $p(x|C)$ and class prior $p(C)$. The question is now how we can write the class likelihood in terms of the distributions specified in the generative model above. As observation $x$ is only dependent on $s$, the class likelihood can be obtained by marginalizing over the intermediate variable $s$:

\begin{align}
    p(x|C) &= \int p(x|s, C) p(s|C) ds \\
         &= \int p(x|s) p(s|C) ds
\end{align}

We can now write the posterior distribution as follows, where all the distributions have been specified:

\begin{align}
    p(C|x) &= \frac{p(x|C) p(C)}{p(x)} \\
           &=\frac{\int p(x|s) p(s|C) ds \ p(C)}{p(x)}.
\end{align}

\paragraph{Choosing an action to minimize cost}
If the task were just a perceptual task where the animals were rewarded equally for reporting the correct class, the optimal Bayesian decision maker should compare the two class posteriors and report the class with higher probability:

\begin{align}\label{eq:simple_d}
    d &= log \frac{p(C = 1 |x)}{p(C = -1 | x)},
\end{align}
where $d$ is the decision variable and its sign indicates the chosen class. However, the key component of the perceptual gambling task was the asymmetric reward cued by flashing lights. According to BDT, the animal's decision should consider the uneven `cost' of the actions and choose the one with minimal cost on each trial. We define the cost function as $\lambda (a | C, v)$, denoting the cost (loss of reward) of action $a$ when the correct class is $C$ and the flashing condition is $v$. Let $v = 1$ denote flashes on the high side, $v = -1$ for flashes on the low side, and $v = 0$ denote no flashes at all. Further, let the base correct reward be $r$ and the reward multiplier be $\kappa$. Then the cost function of the action $a$ given the correct class $C$ and flashing side $v$ can be summarized in the following table: 

\begin{table}
\centering
\begin{tabular}{ |c|ccc|ccc| } 
 \hline
 \multirow{2}{4em}{$\lambda(a | C, v)$} & \multicolumn{3}{|c|}{$C = 1$} & \multicolumn{3}{|c|}{$C = -1$} \\

 & $v = 1$ & $v = -1$ & $v = 0$ & $v = 1$ & $v = -1$ & $v = 0$ \\
 \hline
 $a = 1$ & 0 & 0 & 0 & $r$ & $\kappa r$ & $r$ \\
 \hline
 $a = -1$ & $\kappa r$ & $r$ & $r$ & 0 & 0 & 0 \\
 \hline
\end{tabular}
\vspace{0.1in}
\caption{The action cost table.}
\label{tab:action_cost}
\end{table}

For example, the action cost of reporting class 1 ($a = 1$) when the true class is -1 ($C = -1$) and the flashing side is also -1 ($v = -1$) is $\kappa r$, as the animal is `missing out' on $\kappa r$ reward if it reported correctly. The action cost of reporting class 1 ($a = 1$) when the true class is 1 ($C = 1$) is 0 regardless of the flashing side, as it is the only action to be rewarded in this scenario. To incorporate the cost function into the decision variable, we assume the animal has a representation of the posterior-weighted cost for each action. Concretely, the posterior-weighted cost for choosing class 1 becomes  

\begin{align}\label{eq:d1}
    d_{1} & = \lambda(a = 1 | C = 1, v)^\rho p(C = 1 |x) + \lambda(a = 1 | C = -1, v)^\rho p(C = -1 |x).
\end{align}
And for class -1 becomes

\begin{align}\label{eq:d2}
    d_{-1} & = \lambda(a = -1 | C = 1, v)^\rho p(C = 1 |x) + \lambda(a = -1 | C = -1, v)^\rho p(C = -1 |x),
\end{align}
where $\rho$ is the exponent on the animal's utility function. In our task, we only tested 2 costs, so we could have, instead of an exponent, included a multiplicative reward scaling parameter. However, this is a classic functional form for marginal utility, that allows us to interpret animals with $\rho < 1$ as risk averse and animals with $\rho > 1$ as risk seeking.
Finally, the decision variable can be expressed as the log ratio between $d_{1}$ and $d_{-1}$:

\begin{align}
    d & = log \frac{d_{1}}{d_{-1}} \\
     &= log \frac{\lambda(a = 1 | C = 1, v)^\rho p(C = 1 |x) + \lambda(a = 1 | C = -1, v)^\rho p(C = -1 |x)}
    {\lambda(a = -1 | C = 1, v)^\rho p(C = 1 |x) + \lambda(a = -1 | C = -1, v)^\rho p(C = -1 |x)}.
\end{align}
From the cost function table, we know that $\lambda(a = 1 | C = 1, v) = \lambda(a = -1 | C = -1, v) = 0$, $d$ thus becomes 

\begin{align}\label{eq:d}
    d & = log \frac{\lambda(a = 1| C = -1, v)^\rho p(C = -1 |x)}{\lambda(a = - 1 | C = 1, v)^\rho p (C = 1 | x)} \\ 
    & = log \frac{\lambda(a = 1| C = -1, v)^\rho}{\lambda(a = - 1 | C = 1, v)^\rho} + log \frac{p(C = -1 |x)}{p (C = 1 | x)} \\ 
    & = log \frac{\lambda(a = 1| C = -1, v)^\rho}{\lambda(a = - 1 | C = 1, v)^\rho} - log \frac{p(C = 1 |x)}{p (C = -1 | x)}.
\end{align}

Different from \eqfmt{\ref{eq:simple_d}}, the sign of the decision variable $d$ takes on the opposite value of the final class of choice. The reversion is due to the fact that the goal is to minimize the action cost rather than maximize the posterior distribution. Finally, we converted the decision variable into a probability of choosing $C = 1$ using a \textit{logistic} function:

\begin{align}
    p\mathrm{(Choose \ C=1}|\Theta, s, v) = \frac{1}{1 + e^{d}}
\end{align}
where $\Theta$ refers to all the parameters in the model. 

\paragraph{The three-agent model}
We observed that several animals exhibited `lapses': poor performance even on very easy stimuli \citep{pisupatiLapsesPerceptualJudgments2019}. In order to account for this behavior, we developed a three-agent model that includes a `rational' agent that outputs $p\mathrm{(Choose \ C=1}|\Theta, s, v)$ from the BDT model, and two stimulus-independent agents that either habitually choose the low or high port. The choice on each trial becomes a weighted outcome of the votes from three agents with their respective mixing weights $\omega$, each implementing a different behavioral strategy. Formally, 

\begin{align}
    p(\mathrm{Choose\ C=1} |\Theta, \vec{\omega}, s, v) &= \vec{P} \cdot \vec{\omega} \\
        &= p(\mathrm{Choose \ C=1 | \Theta, s, v}) \cdot \omega_{rational} + 1 \cdot \omega_{high} +
        0 \cdot \omega_{low} \\
    \sum{\vec{\omega}} &= 1
\end{align}

The full model we used to fit animal behavior is thus a BDT-inspired hybrid model, we refer to it as the `mixture-BDT' model. For notation simplicity,  in the following sections we will use $p_{1}$ to denote $p(\mathrm{Choose\ C=1} |\Theta, \vec{\omega}, s, v)$. 

\subsection*{Analysis}
For all analyses, we excluded time out violation trials (where the subjects disengaged from the ports for more than 30 s during the trial). All analysis and statistics were computed in R (version 3.6.3, R Foundation for Statistical Computing, Vienna, Austria). 

\paragraph{Generalized Linear (Mixed-Effects) Models}
Generalized linear models (GLM) and generalized linear mixed-effects models (GLMM) were fit using the \texttt{stats} and \texttt{lme4} R packages \citep{batesFittingLinearMixedEffects2015}. To test whether the animals were sensitive to both tone frequency and flashing side, we specified a mixed-effects model where the probability of choosing the high port was a \textit{logistic} function of $\log_2(kHz)$, the flashing side and their interaction as fixed effects. The flashing side is a categorical variable with three levels: low side flash, high side flash and no flash. The rat and an interaction of rat, $\log_2(kHz)$ and the flashing side are modeled as within-subject random effects. In standard R formula syntax:

\begin{align}
    \mathsf{chose\_high} \sim \mathsf{log2\_kHz * flash\_side + (log2\_kHz * flash\_side | subjid)}
\end{align}
where $\mathsf{chose\_high}$ is 1 if the high port was chosen and 0 if the low port was chosen; $\mathsf{subjid}$ is the subject ID for each rat.

To test whether an individual animal was sensitive to tone frequency and flashing side with each $\sigma_s$ and $\kappa$ combination, we specified a GLM as follows: 

\begin{align}
    \mathsf{chose\_high} \sim \mathsf{log2\_kHz * flash\_side}
\end{align} 
Only the $\sigma_s$ and $\kappa$ combination that resulted in a significant main effect of $\mathsf{flash\_side}$ was included in this study (see \tabfmt{\ref{tab:task_params}}).

To test whether the outcome of the previous trial affected choice on the current trial, we first classified the previous trial's outcome into four categories: the animal chose the low port and was rewarded, the animal chose the low port and was unrewarded, the animal chose the high port and was rewarded, and the animal chose the high port and was unrewarded. A GLMM was specified:

\begin{align}
    \mathsf{chose\_high} \sim \mathsf{log2\_kHz + prev\_outcome\ + (log2\_kHz + prev\_outcome | subjid)}
\end{align}
where $\mathsf{prev\_outcome}$ is a categorical variable with four levels as described above.

To test whether the animal has a tendency to repeat its previous choice, we specified a GLMM as follows:

\begin{align}
    \mathsf{chose\_high} \sim \mathsf{log2\_kHz + prev\_choice\ + (log2\_kHz + prev\_choice | subjid)}
\end{align}
where $\mathsf{prev\_choice}$ is 1 if the high port was chosen on the previous trial and 0 if the low port was chosen.

\paragraph{Trial difficulty analysis}
To understand how perceptual difficulty affected the animal's shift towards the flashing side, we employed a model-based analysis. First, we obtained the animal's perceptual sensitivity $\sigma_p$ using the aforementioned mixture-BDT model. Then, we computed Z-score for each tone frequency $s$ presented to this animal using the formula: $Z = (s - 3) / \sigma_p$. Based on the Z-score, the middle 33\% trials ($- 0.426 \le Z < 0.426$) are labeled as `Hard' trials, the 16.5\% left and right to the hard trials are labeled as `Medium' trials ($-0.95 \le Z < -0.426$; $0.426 \le Z < 0.95$), and the 16.5\% left and right to the medium trials are labeled as `Easy' trials ($Z < -0.95$; $Z \ge 0.95$). By dividing trials this way, we ensured equal proportions of easy, medium and hard trials while taking into account the animal's perceptual sensitivity. After computing the absolute change in percentage choosing the high port induced by light flashes for each difficulty condition, we performed a linear mixed-effects model (LMM) to test significance:

\begin{align}
    \mathsf{delta} \sim \mathsf{difficulty + (difficulty | subjid)}
\end{align}
where $\mathsf{delta}$ refers to the absolute change.

\paragraph{Model fitting}
Following modern statistical convention, we estimated the posterior distribution over model parameters with weakly informative priors using the \texttt{rstan} package (v2.21.2; Stan Development Team, 2020). \texttt{rstan} is the R interface of Stan (Stan Development Team, 2020), a probabilistic programming language that implements Hamiltonian Monte Carlo (HMC) algorithm for Bayesian inference. The prior over the utility exponent $\rho$ was $Lognormal(\log(1), 0.3)$, a weakly informative prior that prefers $\rho$ to be to risk-neutral. The prior over perceptual noise $\sigma_p$ was $Lognormal(\log(0.3), 0.1)$, a reasonable range in $\log_2(kHz)$ space. The prior over the mixing weights $\vec{\omega}$ was a Dirichlet distribution with the concentration parameter $\alpha = [6, 2, 2]$. The resulting $\omega_{rational}$ distribution was broad and had the mean of 0.6, both $\omega_{high}$ and $\omega_{low}$ distribution had the mean of 0.2. By attributing more weight to the rational agent over the habitual agents, the prior reflected our selection of the experimental animals - only animals whose choices depended on the auditory cue were included. Four Markov chains with 1000 samples each were obtained for each model parameter after 1000 warm-up samples. The $\hat{R}$ convergence diagnostic for each parameter was close to 1, indicating the chains mixed well. 

\paragraph{Sigmoid function}
The four-parameter sigmoid function was specified as follows:

\begin{align} \label{eq:sigmoid}
    y = w_2(1 - w_1) + \frac{w_1}{1 + e^{-b(x - x_0)}}
\end{align} 
where x is the tone frequency in , y is the probability choosing high port, and the four parameters are: $x_0$, the inflection point of the sigmoid, controlling horizontal shifts; $b$, the slope of the sigmoid; $1 - w_1$, the total lapse rate, and $w_2$, representing the fraction of lapses that are low to high lapses. The sigmoid model was fit individually to each flash condition in each subject's dataset using Stan.

\paragraph{Synthetic datasets}
To test the validity of the mixture-BDT model, we first created synthetic datasets with parameters generated from the prior distributions described above. The model was used to fit on the synthetic datasets, and was able to recover the generative parameters accurately (\figfmt{\ref{supp:sanity}}). This assured that the model had no systematic bias in estimating the parameters.

\paragraph{Model prediction confidence intervals}
To estimate the confidence intervals with model prediction as in \figfmt{\ref{fig:model}A}, we first generated a synthetic dataset with regularly spaced sound frequencies (incremented by 0.01). After parameter sampling in each iteration (in the \texttt{generated quantities} block), the sampled parameters were used to predict the choices given the synthetic offers. The resulting output is a \texttt{n\_iter $\times$ n\_sound} matrix, where \texttt{n\_iter} is the number of iterations and \texttt{n\_sound} is the length of unique stimulus frequencies. Finally, $\pm 80$, $\pm 95$ and $\pm 99$ confidence intervals for each offer were estimated by taking the respective quantiles of the \texttt{n\_iter} predicted choices. 

\paragraph{Mixture-BDT optimality analysis}
To understand the relationship between $\rho$ and $\sigma_p$ in obtaining maximum possible reward, we first created a synthetic task dataset with 1000 trials. The tone frequency was drawn from a truncated Gaussian centered at 3 and a standard deviation of 0.6. The $\kappa$ was set to 5. There were equal proportions of the high flash, low flash and no flash trials ($\sim 33 \%$). We then created a mixture-BDT agent that is fully rational ($\vec{\omega} = [1, 0, 0]$). A grid search was performed to find the total reward for each combination of $\rho$ (0 to 1.5, incremented by 0.1) and $\sigma_p$ (0 to 0.5, incremented by 0.1). The total reward ($R$) was computed as follows:

\begin{align}
    R = \sum_{i}^{1000} (1 - p_{1}^i) \cdot r_{low}^{i} + p_{1}^i \cdot r_{high}^{i}
\end{align}
where $p_{1}^i$ is the probability of choosing the high port from the mixture-BDT agent on the i-th trial, $r_{low}^{i}$ is the reward delivered if choosing the low port on the i-th trial, and $r_{high}^{i}$ is the reward delivered if choosing the high port. 

\paragraph{Code availability}
The perceptual gambling dataset from the 7 animals and all the code used for analyses are freely available online at \texttt{https://github.com/erlichlab/perceptual-gambling}

\section*{Results}

\subsection*{The perceptual gambling task}
\begin{figure*}[htbp]
\includegraphics[width=1\linewidth]{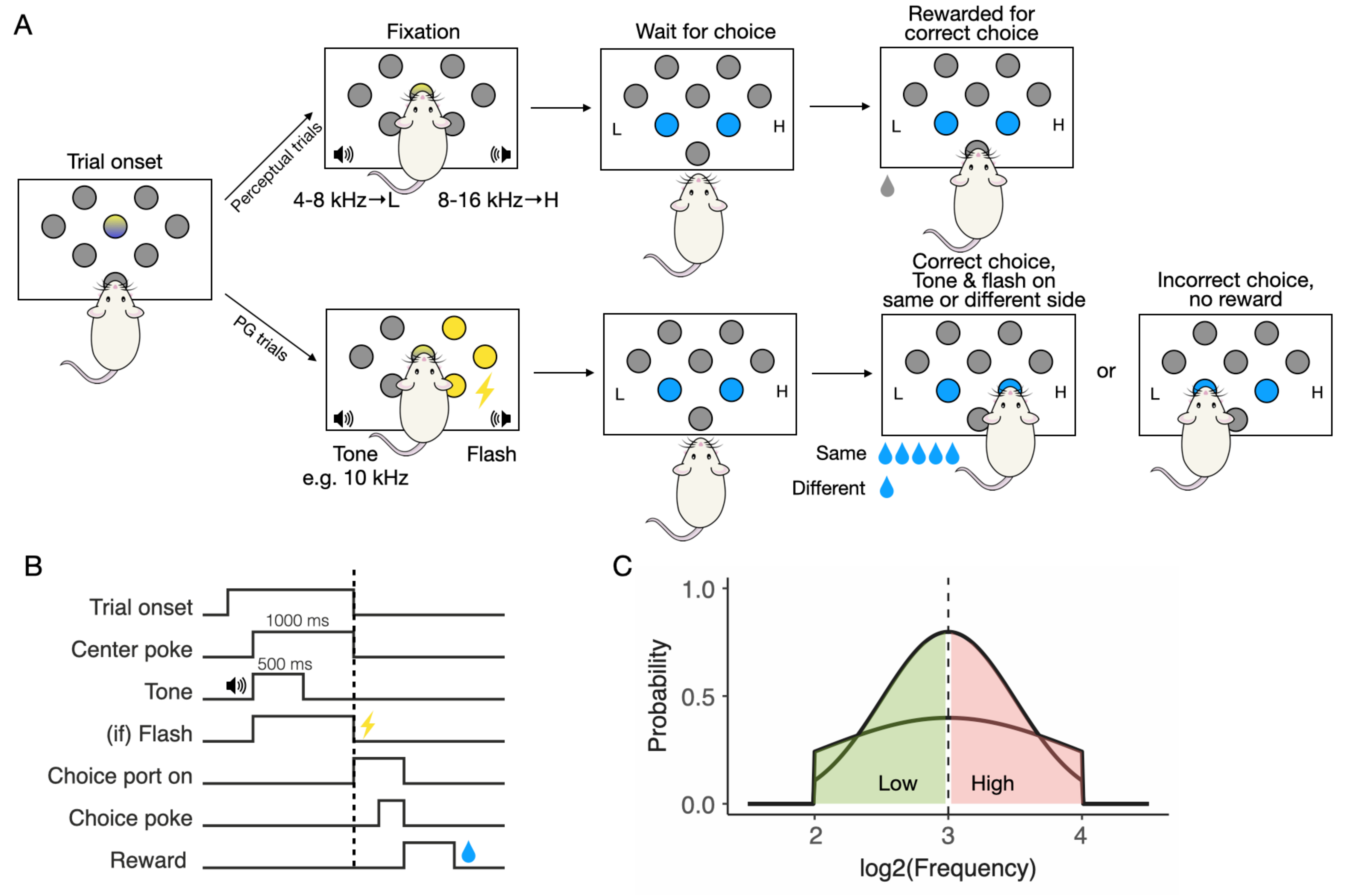}
\caption{The perceptual gambling task.
\textbf{A.} Schematic of the perceptual gambling task. 
\textbf{B.} Schematic of the trial structure. 
\textbf{C.} The tone frequency was drawn from a truncated Gaussian, $\psi (3, \sigma_s, 2, 4; s)$, where $\sigma_{s}$ was tuned for each animal in different training stages, but was otherwise fixed within sessions. A larger $\sigma_s$ results in a wider distribution and easier trials, and a smaller $\sigma_s$ results in a distribution more concentrated around the decision boundary.
}
\label{fig:task}
\end{figure*}

To establish a rodent framework to study decisions guided by both perceptual and value cues, we developed the perceptual gambling task. It was named so because although the correct decision was only informed by the perceptual cue, a reward-maximizing subject would choose the side with larger reward when the perceptual evidence was weak, effectively `gambling' for more reward (\figfmt{\ref{fig:task}A}). For example, imagine the subject was 75\% certain that the stimulus should be categorized as `high'. But, on this trial, the subject knew that a correct `high' response would be rewarded with 1 drop of water, while a correct `low' response would fetch 8 drops. Then, the expected value of responding high would be $P_{high}\cdot V_{high} = 0.75 \cdot 1 = 0.75$. The expected value of responding low would be $P_{low}\cdot V_{low} = (1 - 0.75) \cdot 8 = 2$. As such, the task invites the animal to gamble based on its perceptual confidence, which was experimentally varied by requiring subjects to make easy and difficult perceptual decisions, and the values of the two responses. 

Subjects were first trained on the pure perceptual version of the task with symmetric rewards, and we refer to these trials as `perceptual' trials. On each trial, after self-initiation by poking into the center port, subjects fixated for 1 s while a tone would play from both speakers. Its frequency relative to 8 kHz indicated whether the left or right port was correct (counter-balanced across animals). We refer the correct port for frequencies lower than 8 kHz as `low port', and the correct port for frequencies higher than 8 kHz as `high port'. The tone frequency (in $\log_2(kHz)$ space) was drawn from a truncated Gaussian distribution $\psi (3, \sigma_s, 2, 4; s)$, where $\sigma_s$ is the standard deviation and controls the difficulty of the perceptual task. The smaller $\sigma_s$ is, the more concentrated the auditory stimulus is around the decision boundary, and the more perceptually challenging the trials will be (\figfmt{\ref{fig:task}C}). Once the animals showed good performance on the perceptual trials, we introduced the value cues during fixation by flashing the yellow LEDs of the three left or right ports (\figfmt{\ref{fig:task}A}). The choice of the flashing side was independent of the tone frequency. We delivered the perceptual and value cues through the auditory and visual modality, respectively, to avoid any effects from intra-modality attention. For example, a louder tone from one side might interact with the animal's judgment of its frequency. If the correct port was on the same side as the flashing ports, correct responses resulted in a large reward (base reward $\times$ $\kappa$). Alternatively, if the flashing side was on the incorrect side, the animal was only rewarded the base amount for choosing the correct port. These trials are referred as `perceptual gambling (PG)' trials. The perceptual trials were randomly interleaved with PG trials in a session, the ratio of the two trial types was different for each animal. 

Training of the task was difficult, as the animal's performance was highly sensitive to the values of task parameters, especially the difficulty of the perceptual task ($\sigma_s$) and reward asymmetry ($\kappa$). This is not surprising, \citet{kepecsNeuralCorrelatesComputation2008a} trained rats on an odor discrimination task and the animals were only rewarded (nor not) after a variable delay. While it was waiting for the reward, the rat had an option to `re-initiate' the trial by leaving the choice port and start again, the frequency of which should correlate with its confidence of the perceptual decision. It was later reported that the training was also very parameter-sensitive, as the reward delay interacted with the rat's temporal discounting function, similar to how the perceptual difficulty interacted with the reward sensitivity in our task \citep{kepecsComputationalFrameworkStudy2012}. Nonetheless, we successfully trained 7 animals on the task (see \tabfmt{\ref{tab:task_params}} for the effective task parameters for each animal). These 7 subjects were sensitive to both the auditory cue and the flashes with at least one set of task parameters, quantified by a generalized linear model (GLM). When the flash was first introduced, the animals did not show any bias towards the flashing side. Thus, the shifts in choices caused by the flashes were induced by learning, rather bottom-up attention.

\begin{table}[!ht]
 \centering
 \begin{tabular}{|c|c|c|c|c|c|c|c|}
 \hline
  & 2077 & 2078 & 2083 & 2085 & 2109 & 2124 & 2143  \\ \hline
  $\sigma_s$ & 1 & 1 & 1 & 1 & 0.3 & 0.3 & 0.3  \\ 
$\kappa$ & 5 & 5 & 5 & 5 & 25 & 20 & 15 \\ 
 \hline
 \end{tabular}
 \vspace{0.1in}
 \caption{The perceptual difficulty ($\sigma_s$) in $\log_2(kHz)$ space and reward multiplier of the flashing side ($\kappa$) effective for each animal.}
\label{tab:task_params}
\end{table}

\subsection*{Animal behavior}
\begin{figure*}[htbp]
\includegraphics[width=1\linewidth]{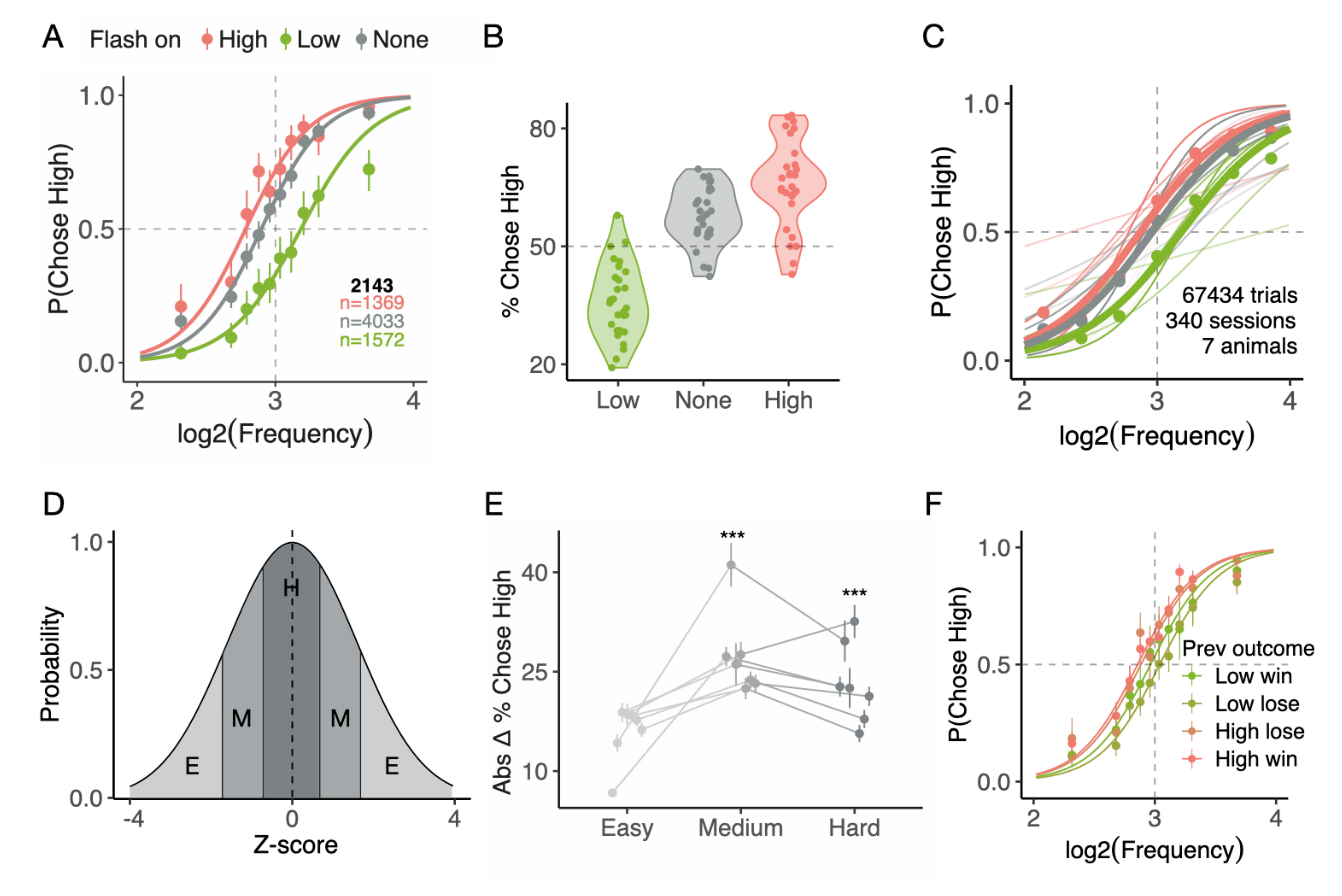}
\caption{Animal behavior in the perceptual gambling task.
\textbf{A.} Example subject performance from 28 sessions with $\sigma_s$ = 0.3 and $\kappa$ = 15. The probability of choosing the port associated with high frequencies is plotted as as a function of $\log_2(kHz)$, where 3 was the decision boundary. The circles with error bars are the mean $\pm 95$\% binomial confidence intervals. The lines are the psychometric curves generated by a GLM. The colors represent the flashing condition, with red - high side flash, green - low side flash, gray - no flash. 
\textbf{B.} Violin plot showing distribution of the percentage choosing the high port under three flashing conditions. Data used here is the same as in A. Each dot represents the mean percentage choosing the high port from this condition in each session. 
\textbf{C.} Performance from all 7 subjects trained on the task. The circles with error bars are the mean $\pm 95$\% binomial confidence intervals using the population data. The thick colored lines are the psychometric curves generated by a generalized linear model fit to all the sessions together. The thin colored lines are from a generalized linear model fit to each subject's dataset individually.
\textbf{D.} Trials were divided into hard (H, hardest 33\% of trials), medium (M, next hardest 33\% of trials) and easy (E, easiest 33\% of trials) based on each animal's perceptual noise. See details in Methods. 
\textbf{E.} The absolute change in percentage choosing the high port induced by light flashes in easy, medium and hard trials. The changes induced by either low and high flashes were averaged. Animals significantly shifted more in medium than easy trials ($\beta_{medium} = 0.11 \pm 0.03, p < 0.001$), and more in hard than easy trials ($\beta_{hard} = 0.07 \pm 0.03, p < 0.001$).  
\textbf{F.} 2143's choice on the current trial was influenced by the outcome of the previous trial. 
}
\label{fig:behavior}
\end{figure*}

We trained 7 male rats (4 Brown Norway, 3 Sprague Dawley) on the perceptual gambling task. As expected, the animals' performance was a function of evidence strength (see an example animal in \figfmt{\ref{fig:behavior}A}; see population aggregates in \figfmt{\ref{fig:behavior}C}). Moreover, the animals reliably shifted their choices towards the side with flashing lights (see example shift in \figfmt{\ref{fig:behavior}B}, all animals in \figfmt{\ref{supp:sanity}}). These effects were quantified using a  generalized-linear mixed-effects model (GLMM). There was a significant main effect of tone frequency ($\beta_{freq} = 2.86 \pm 0.04, p < 0.001$), and a significant interaction between tone frequency and flash on the low side ($\beta_{freq:side} = -0.13 \pm 0.15, p < 0.05$). Interestingly, flashing on the high side did not affect behavior significantly on the group level ($p > 0.05$). 

The premise of the task is that the animal should prefer the flashing side more when the stimulus is closer to the decision boundary. In other words, when perceptual evidence is weak, value information should have more influence on choice. To test whether this was true, we divided trials into easy, medium and hard trials based on their evidence strength relative to the animal's perceptual noise, which was estimated using a Bayesian model (see description in the next section; \figfmt{\ref{fig:behavior}D}).
With a linear mixed-effects model (LMM), we found that trial difficulty significantly affected the absolute shift in percentage choosing the high port ($\beta_{hard} = 0.07 \pm 0.03, p < 0.001$; $\beta_{medium} = 0.11 \pm 0.03, p < 0.001$, \figfmt{\ref{fig:behavior}E}). 
This is in line with our prediction that the subjects would shift more in medium and hard than easy trials, although the reason why they shifted more for medium than hard trials was unclear. 
Finally, we found that the choices on the current trial were significantly influenced by the outcome of the previous trial (GLMM, all $p < 0.001$; see an example in \figfmt{\ref{fig:behavior}F}, see all animals in \figfmt{\ref{supp:history}}). Although the individual history effects differ, overall, the animals had a tendency to repeat its previous choice ($\beta_{prev\_choice} = 0.60 \pm 0.05, p < 0.001$). 

\subsection*{A three-agent mixture model with Bayesian decision theory}
While the GLMM results indicated that the animal's choices were sensitive to both perceptual and value cues, it does not provide insight into the cognitive processes underlying task performance. To better understand how our animals integrated perceptual and value information, we developed a Bayesian decision theory (BDT) model \citep[following][]{maBayesianDecisionModels2019}. Bayesian modeling starts with a generative model, specifying how the subject's observation come about given the statistics of the environment, which is usually set by the experimenters. Using the Bayes' rule, the subject then combines its prior with the observation to obtain the posterior, a probability distribution reflecting both the observed measurement and its prior belief. Finally, a Bayesian decision maker chooses an action in a principled manner by minimizing a cost function $C(s, a)$, which is determined by the state of the world $s$ and the action $a$. The BDT framework is well suited for our task, as the animal acts by integrating a noisy perceptual stimulus (observation) and asymmetric reward associated with each choice (cost function).

\begin{figure*}[htbp]
\includegraphics[width=1\linewidth]{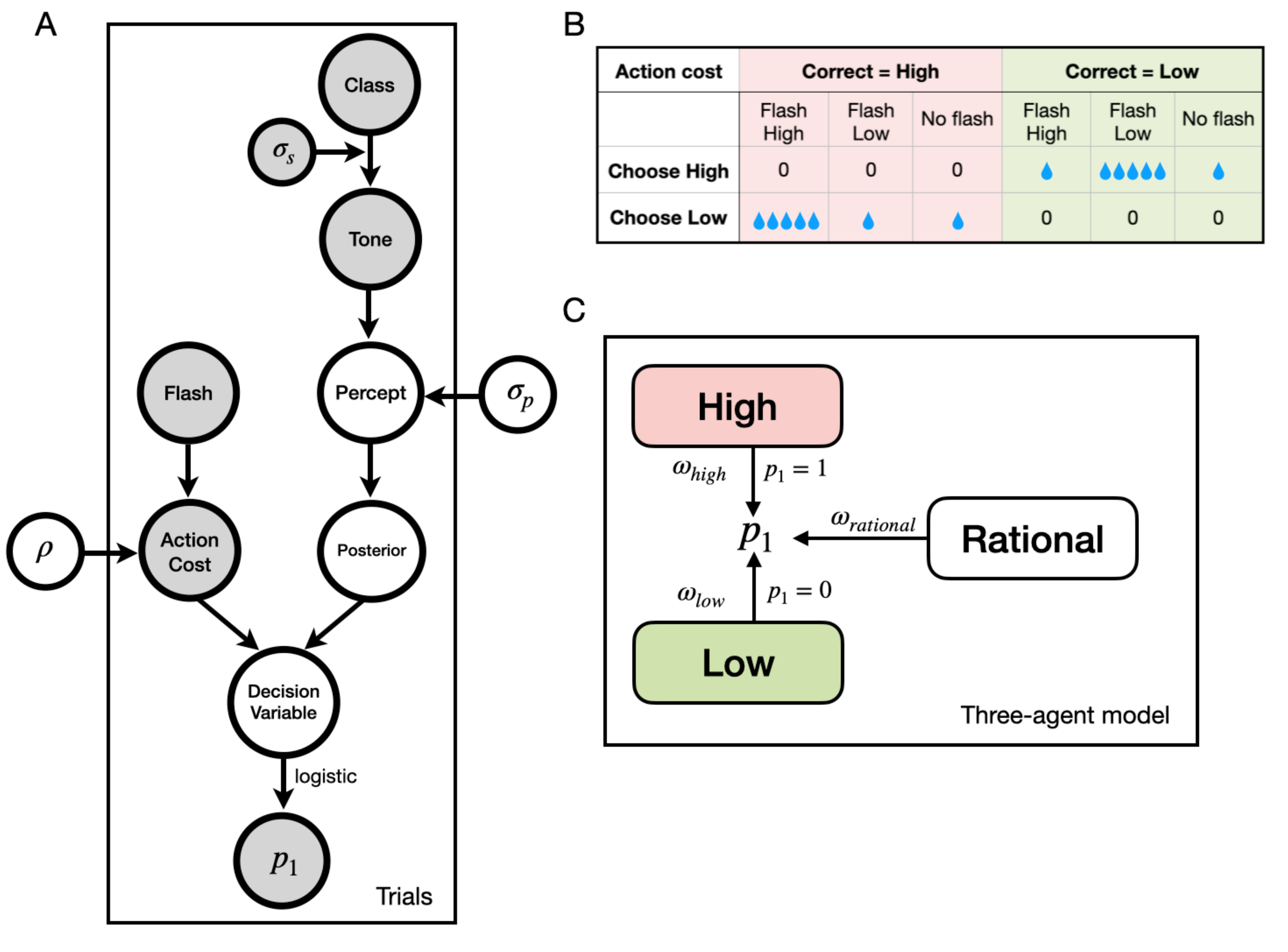}
\caption{The mixture-BDT model.
\textbf{A.} Graphical representation of just the BDT model. Using the plate notation, the variables shaded in gray are known or observable to the experimenters, and the variables in white are latent variables. The arrows indicate dependency between the variables. The two parameters to estimated are $\rho$, the curvature of utility function, and $\sigma_p$, the perceptual noise. See Methods for details.
\textbf{B.} The cost function table. The cells where the correct choice is `high' are shaded in red, and the correct choice is `low' are shaded in green. Action cost is defined as the cost, or missed reward, from choosing the incorrect port. Naturally, it is 0 when the animal chooses correctly. It is the $base~reward$ (denoted by a single water drop) when the animal chooses incorrectly and the flash is not on the other side. It is $base~reward \times \kappa$ (denoted by multiple water drops) when the animal chooses incorrectly and the flash is on the other side. 
\textbf{C.} The three-agent mixture model. The animal's final choice is modeled as a weighted average of the three agents, each implementing a different behavioral strategy to perform the task. The `rational' agent outputs the probability of choosing high from the BDT model; the `high' agent always chooses the high port ($p_1 = 1$) and the `low' agent always chooses the low port ($p_1 = 0$). The final probability of choosing high port ($p_1$) is a weighted sum with the agent's respective weight $\vec{\omega}$, where $\sum{\vec{\omega}} = 1$.
}
\label{fig:bdt}
\end{figure*}

Next, we will briefly describe the model (\figfmt{\ref{fig:bdt}A}, see modeling details in Methods). We start with the generative model, which specifies how the subject makes an observation $x$ given the stimulus frequency $s$ on each trial (\figfmt{\ref{fig:method}A}). Recall that the stimulus was drawn from a truncated Gaussian centered at $3$ with standard deviation $\sigma_s$, which was set by the experimenter. Bayesian models assume that through experience, subjects learn this distribution, and utilize it when inferring the correct class (low or high) given the tone on each trial. Thus, the observation distribution is dependent on two parameters: $\sigma_s$, which is known, and $\sigma_p$ as in $p(x|s) \sim \mathcal{N}(s, \sigma_p)$, denoting the perceptual noise of each animal.  
The observation is then combined with class prior ($p = 0.5$ for each class) to compute the class posterior, representing the animal's belief of each class given just the perceptual cue. To incorporate value information, we constructed a cost function where the choice is mapped to an action cost under different flash conditions (\figfmt{\ref{fig:bdt}B}, \tabfmt{\ref{tab:action_cost}}). For example, when the high side is flashing and the correct class is high and the animal chooses low, the action cost would be $base~reward \times \kappa$, a miss of considerable size. We included an additional parameter $\rho$ as the exponent on the action cost, which is equivalent to the curvature of the animal's utility function: $U = V^\rho$, where $U$ denotes utility and $V$ is value. Finally, the $\rho$-adjusted action cost of choosing each class is integrated with its class posterior as the decision variable (\eqfmt{\ref{eq:d}}), which is transformed into a probability of choosing the high port using a \textit{logistic} function. 

We observed that some animals exhibited a constant, stimulus-independent rate of error known as `lapse'. Recently, it has been suggested to reflect exploration in a changing environment \citep{pisupatiLapsesPerceptualJudgments2019}. To account for the lapses, we developed a `three-agent' model that includes a `rational' agent that outputs the probability of choosing the high port from the BDT model, a habitual `high' agent that always chooses the high port, and a `low' agent that always chooses the low port (\figfmt{\ref{fig:bdt}C}). The choice on each trial is thus a weighted outcome of the votes from three agents with their respective mixing weights $\omega$, each implementing a different behavioral strategy. We refer to the final hybrid model as the `mixture-BDT' model.  

\subsection*{The mixture-BDT model is insufficient to account for subjects' behavior}
\begin{figure*}[htbp]
\includegraphics[width=1\linewidth]{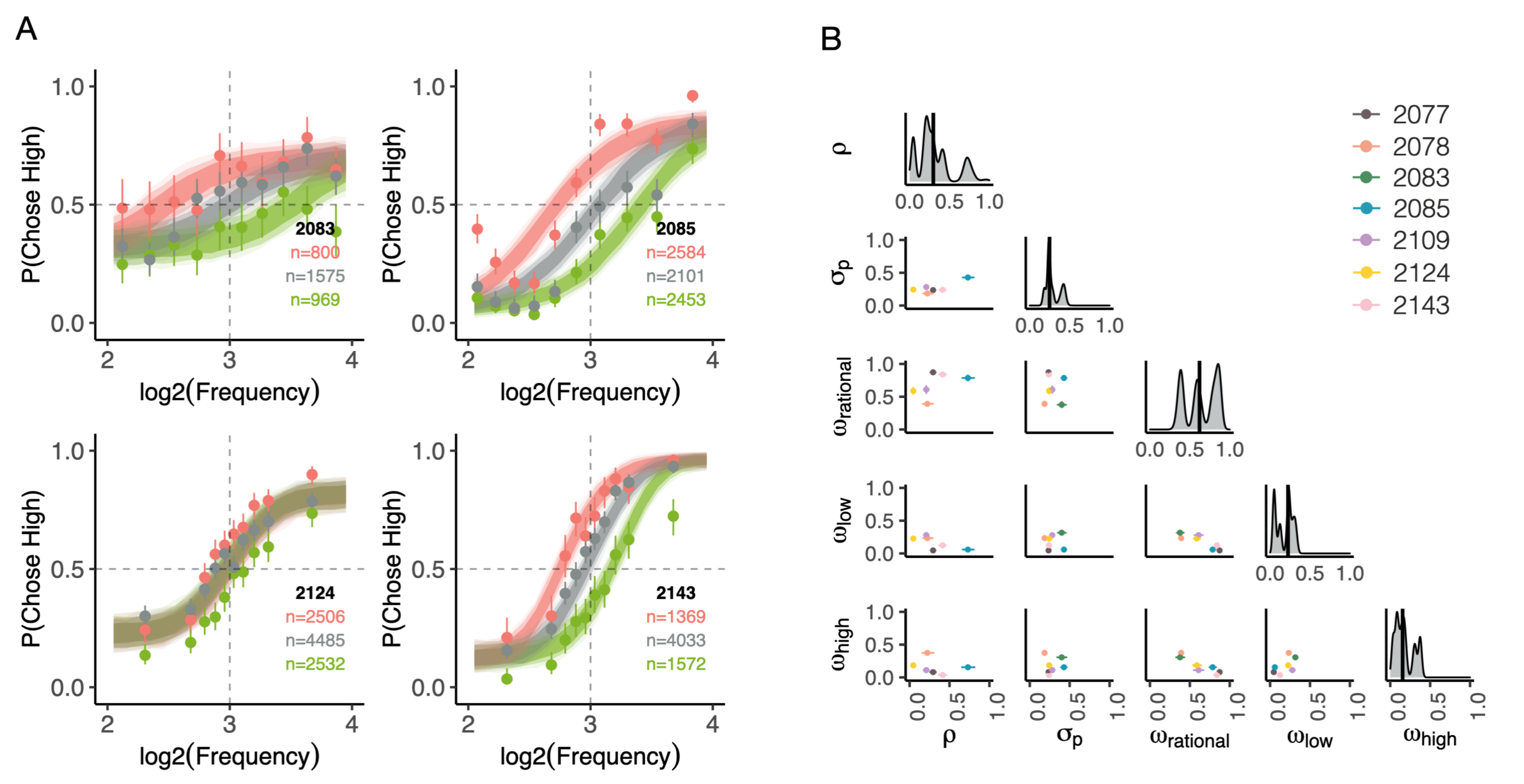}
\caption{The mixture-BDT model fits.
\textbf{A.} Example predictions using the model estimates. The circles with error bars are the binned mean $\pm 95$\% binomial confidence intervals from data. The colors represent the flashing condition. The ribbons are model predictions generated using the fitted parameters. The dark, medium and light shade represent $\pm 80$\%, $\pm 95$\% and $\pm 99$\% confidence intervals, respectively. The model fit some animals fine (e.g. 2085, 2143) but was not able to fit other animals exhibiting vertical shifts due to flashes (e.g. 2124). 
\textbf{B.} Summary of the parameters of 7 animals. The mean and $\pm 90$\% confidence interval of each parameter pair are shown in the off-diagonal, colored by subject. Density plots of all fit posterior samples (n = $4000 \times 7$) for each parameter are on the diagonal, the black bar is the median.
}
\label{fig:model}
\end{figure*}

We first validated that the model can correctly recover generative parameters from synthetic data (\figfmt{\ref{supp:sanity}}). We estimated the joint posterior over the parameters for each animal separately using Hamiltonian Monte Carlo sampling in Stan (see example animals in \figfmt{\ref{fig:model}A}, see all animals in \figfmt{\ref{supp:bdt_fits}}). Details of the modeling, including the priors, can be found in the Methods section. Overall, the animals all had a concave utility function ($\rho$ = 0.30 [0.04 1.39], median and 95\% C.I. of concatenated posteriors across animals). They had medium to low levels of perceptual noise ($\sigma_p$ = 0.25 [0.17 0.45]), indicating that on average, they were sensitive to tone frequencies roughly 1.18 kHz apart. Consistent with GLMM results, animals with a sharper psychometric curve (e.g. 2143, pink dot in \figfmt{\ref{fig:model}B}) had a smaller $\sigma_p$ than animals with a flatter psychometric curve (e.g. 2083, green dot in \figfmt{\ref{fig:model}B}). 2077, 2085 and 2143 were guided mostly by the rational agent ($\omega_{rational}$ = 0.84 [0.75 0.88], $\omega_{low}$ = 0.06 [0.04 0.14], $\omega_{high}$ = 0.08 [0.02 0.17], concatenated posteriors across these animals). In contrast, 2078, 2083, 2109 and 2124 displayed high levels of stimulus-independent bias ($\omega_{rational}$ = 0.49 [0.34 0.66], $\omega_{low}$ = 0.25 [0.21 0.34], $\omega_{high}$ = 0.24 [0.09 0.39]). However, this model failed to account for several aspect of animals' behavior. First, there is only one parameter, $\rho$, modulating how much subjects shift responses on PG trials, so the model predicts that the flash-induced shift should be symmetrical for left-flash and right-flash trials, which is not the case in our data. Second, the model predicts that flashes should result in horizontal shifts in the psychometric curve: the shift should depend on the perceptual uncertainty (\figfmt{\ref{supp:bdt_shift}}). In our data, some subjects shifted vertically (\figfmt{\ref{fig:model}A, 2124}): a stimulus independent shift. 

\begin{figure*}[htbp]
\includegraphics[width=1\linewidth]{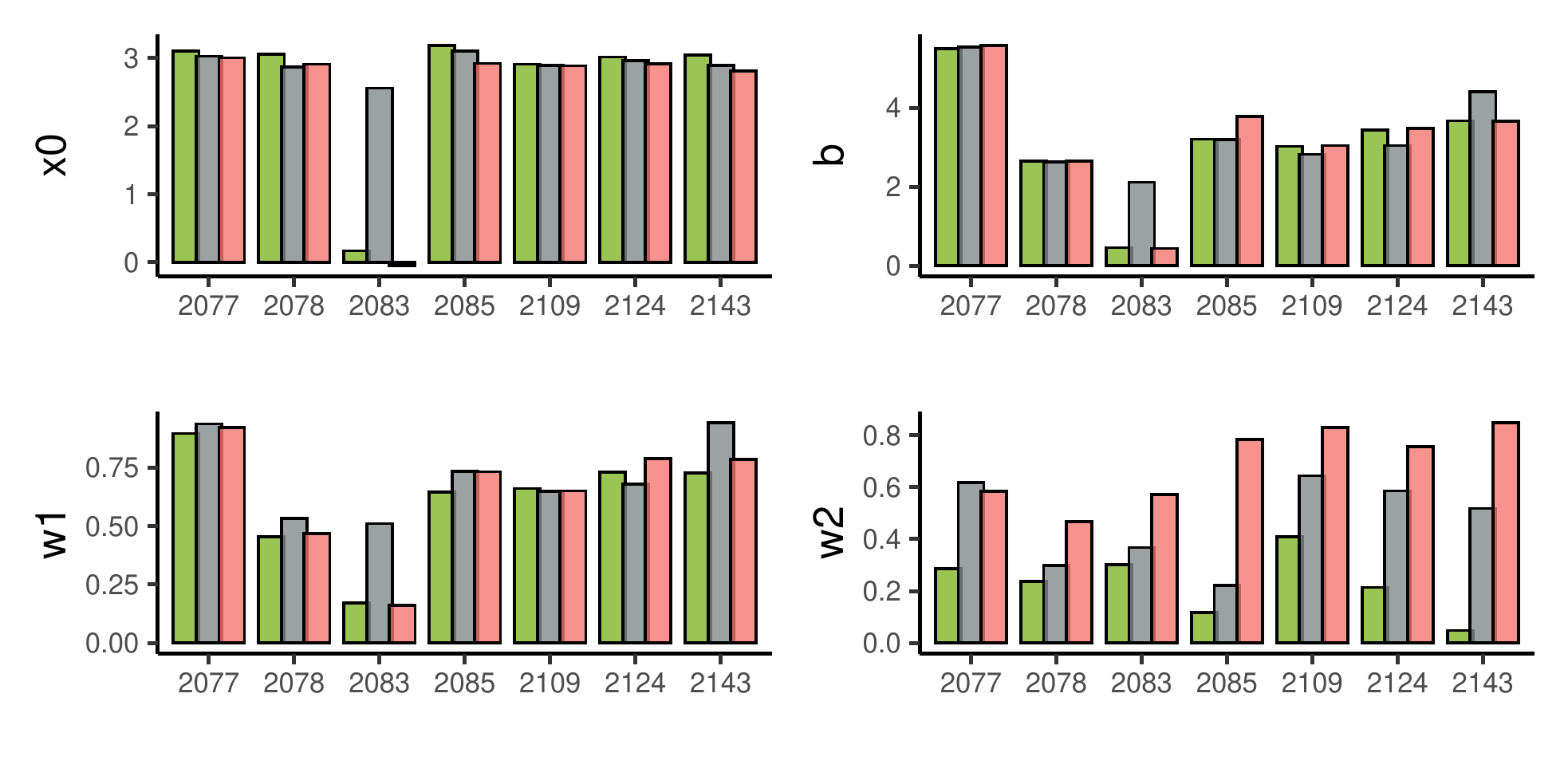}
\caption{The four-parameter sigmoid model fits, shown in maximum a posteriori estimates. The model was fit separately on data from each condition: red - high side flash, green - low side flash, gray - no flash. 
}
\label{fig:sigmoid}
\end{figure*}

To quantify the degree to which the mixture-BDT model failed to fit the data, we refit the data, treating the perceptual trials, left flash and right flash trials as separate datasets, and fit each with a four-parameter sigmoid function (\eqfmt{\ref{eq:sigmoid}}, see details in Methods). 
If, as the mixture-BDT model predicts, flashes induces horizontal shifts (and small increases in slope), then the intercept term, $x_0$, would change the most, with small changes in the slope, $b$. However, in most animals, $w_1$ or $w_2$ changed in the flash trials relative to the perceptual trials, indicating vertical rather than horizontal shifts (\figfmt{\ref{fig:sigmoid}}). Taken together, the modeling result suggests that the animal behavior is not well described by the normative BDT model, even after taking into account lapse, and thus the animals were not optimally integrating the cues. 


\subsection*{The animals are not reward-maximizing}
\begin{figure*}[htbp]
\includegraphics[width=0.8\linewidth]{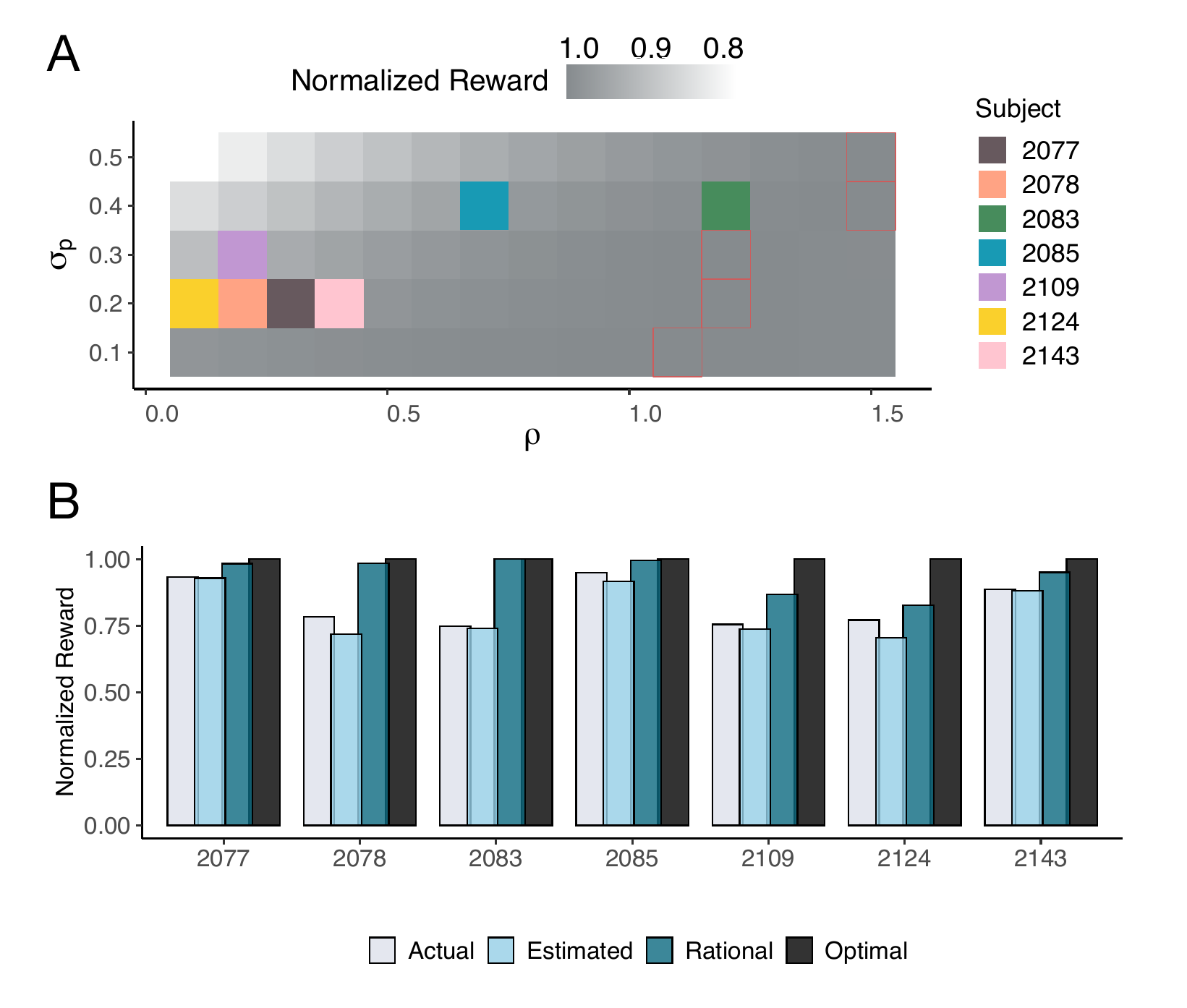}
\caption{The animals were sub-optimal in obtaining reward. 
\textbf{A.} Different values of $\rho$ is required for different $\sigma_p$ to obtain maximum reward. For each $\rho$ and $\sigma_p$ combination, we simulated a rational BDT agent ($\vec{\omega} = [1, 0, 0]$) and computed the reward it obtained using the same task dataset. The color of the heatmap corresponds to the total reward obtained by this parameter combination, normalized  across rows. The MAP estimates of $\rho$ and $\sigma_p$ for each subject with the three-agent BDT model are shown in colored blocks. The $\rho$ that resulted in maximum reward given each $\sigma_p$ is highlighted in red squares. 
\textbf{B.} Comparing reward obtained using four different agents. Actual: the sum of actual reward obtained for this animal. Estimated: the total reward obtained by a BDT agent simulated with fitted $\rho$, $\sigma_p$, $\vec{\omega}$. Rational: the total reward obtained by a BDT agent simulated with fitted $\rho$, $\sigma_p$ and $\vec{\omega} = [1, 0, 0]$. Optimal: the total reward obtained by a BDT agent simulated with fitted $\sigma_p$, the best $\rho$ from \textbf{A} given $\sigma_p$, and $\vec{\omega} = [1, 0, 0]$. All normalized by the maximum reward, which was obtained by the optimal agent. See details in Methods.}
\label{fig:optimality}
\end{figure*}

The vertical shifts induced by the flashes are sub-optimal. How much reward were the subjects missing out on? To answer this, we compared the rewards obtained by the animals with that by an optimal, reward-maximizing Bayesian decision-maker. We define `optimality' here as obtaining the maximum possible reward given a fixed perceptual noise $\sigma_p$ in a particular task dataset. Interestingly, it was found that in order for a purely rational BDT agent to obtain maximum reward, its utility curvature $\rho$ needs to balance with its perceptual noise $\sigma_p$ (\figfmt{\ref{fig:optimality}A}). This can be intuitively understood by going through some examples. For a subject with large $\sigma_p$, it will have a high error rate due to poor perceptual judgment, the strategy to maximize reward would be to choose the flashing side as much as possible and result in a convex utility function. Alternatively, a subject with very small $\sigma_p$ will get most trials correct anyway, it does not need to `value' the flash more than what it represents, resulting in a close-to-linear utility function. Another interesting result from the simulation analysis is that $\rho$ does not affect the total reward much when $\sigma_p$ is small, but its value plays a big role when $\sigma_p$ is large. This in part, explains why subjects like 2077 and 2143 are closest to the `optimal', reward-maximizing agent even with $\rho$ smaller than the best $\rho$ (\figfmt{\ref{fig:optimality}B}). For subjects 2078, 2109 and 2124, their large stimulus-independent bias seemed to be the culprit for obtaining lower reward overall. A fascinating example is 2083, its estimated $\rho$ and $\sigma_p$ combination are close to being optimal (\figfmt{\ref{fig:optimality}A, green square}). The fact that it only obtained 75\% of the maximum reward was entirely due to its lapse rate, as a rational agent with its $\rho$ and $\sigma_p$ obtained just as much reward as the optimal agent. On average, the animals obtained 83.2 $\pm$ 3.3 \% of the maximum reward obtained by their respective optimal agent, modeled with the same perceptual noise. Taken together, the analysis showed that the animals are not optimal in reward-maximization in the task, and this is due to a combination of high lapse rate and extreme risk-aversion. 

\section*{Discussion}
Decision-making is a term referring to the integration and transformation of external information with internal beliefs into an action. The external information may contain perceptual as well as value aspects of the decision required at hand. Despite its importance, only few studies have examined the behavioral and neurobiological underpinnings of decisions that involve percept-value integration. Here, we developed the perceptual gambling task where the rat made choices informed by both perceptual (tone frequency) and value (light flash) cues, on a trial-by-trial basis. Although the subjects did not, on average, shift their choices symmetrically by value as in monkeys \citep{rorieIntegrationSensoryReward2010}, the animals nevertheless showed sensitivity to flashes. We characterized behavior using the Bayesian decision theory, which assumes an optimal integration of individual perceptual noise and reward sensitivity, as well as the statistics of the task environment. It was found that the behavior was not well fit by the BDT model, even after accounting for lapses, because subjects responded to the flashes by shifting their choices in a stimulus independent way. Finally, we quantified the fraction of the reward that animals were foregoing by using a sub-optimal strategy.  Model-based analysis revealed that the missing reward was due to a large lapse rates and risk aversion (extremely concave utility functions). 

Overall, the results show that the animals are not behaving optimally in the task. Their choices were influenced by the previous trial's outcome, even while the tone frequency and flash was independent across trials. \citet{lakDopaminergicPrefrontalBasis2020} also observed strong history effects, but in that task the animal was encouraged to incorporate history information due to the block design. Although Bayesian decision theory has had success in explaining and predicting human behavior \citep[e.g.][]{cogleyAnticipatedUtilityRational2008, kordingBayesianDecisionTheory2006}, it has rarely been used to quantify rodent behavior. One key assumption of the Bayesian decision models is that the subject has \textit{learned} the distributions of latent variables in the task environment and utilized them fully when making inference. It is likely that both the robust history effect and poor fit of the mixture-BDT are a result of incomplete or ongoing learning. For most animals, the training process involved adjusting perceptual difficulty $\sigma_s$ and reward multiplier $\kappa$ periodically to induce significant shifting. As a result, the animals may have internalized the environmental volatility and were actively exploring the reward contingencies. This may also underlie the substantial lapse rate observed in some animals \citep{pisupatiLapsesPerceptualJudgments2019}. Thus, we emphasize that the poor performance of BDT does not suggest that our rats are non-Bayesian agents, it may merely reflect the lack of learning aspects in the theory. In fact, it seems that the monkey performance in a similar task from \citet{rorieIntegrationSensoryReward2010} can be well fit by the BDT, as their value-induced shift in the psychometric curves was horizontal.

The main takeaway from the training process was that task parameters like perceptual difficulty and reward multiplier heavily interacted with the animal's sensory noise $\sigma_p$ and utility exponent $\rho$. Future researchers interested in adopting this framework are encouraged to use a model-based training method. Specifically, when $\sigma_p$ is low and $\rho$ is small, to induce a behavioral shift, the experimenter needs to increase perceptual difficulty and increase the reward multiplier. When $\sigma_p$ is high and $\rho$ is large, to prevent the animal from simply choosing the side with flash and ignoring the sound, the experimenter can reduce perceptual difficulty and decrease the reward multiplier. The animal's $\sigma_p$ can be estimated from the performance on the perceptual trials alone, prior to any value training. Estimating $\rho$ is challenging without training the animal on a task that exposes its subjective utility function. Nonetheless, results from choice under risk showed that most rats have concave utility functions \citep[$\rho$ < 1,][]{zhuFrontalNotParietal2021, constantinopleAnalysisDecisionRisk2019}. On that account, it is reasonable to assume a small $\rho$ in the beginning of training unless the animal showed otherwise. However, we do not exclude the possibility that the animal may `adapt' the shape of its utility function in different contexts, for example, a noisy perceptual decision-maker may deliberately become more risk-seeking to harvest more reward. There is some mixed evidence from behavioral ecology experiments supporting a context-dependent change in utility concavity \citep{kacelnikTriumphsTrialsRisk2013a}. Finally, given the small $\sigma_p$ found in most animals in this study, it is advisable to use a narrower range of auditory stimuli (e.g. 5.65 - 11.31 kHz) to facilitate the training process. 
 
In this manuscript, we did not explicitly test whether the flash-induced shift was due to a perceptual bias or a response bias, which predicts that the value information exerts influence \textit{on} or \textit{separate\ from} sensory processing, respectively. Nonetheless, the perceptual gambling task along with the mixture-BDT model together, can generate specific hypotheses on how to distinguish these two scenarios. Future researchers can record activity from the secondary motor regions and associative sensory regions in well-trained rats to establish correlative relationships. A response bias would predict increased activity in the motor region when the chosen side is cued for higher reward, whereas a perceptual bias would predict differential activity elicited by the same auditory stimulus in the sensory areas under flash and no-flash conditions. Furthermore, causal evidence for either scenario can be obtained with pharmacological and optogenetic inactivations. If by inactivating the secondary motor region the animal simply shifts less to the flashing side, which is equivalent to a decreased utility exponent $\rho$ in the model, the response bias hypothesis will be supported. Even more interestingly, this would suggest a dissociable process of perceptual decision and value computation. On the other hand, the perceptual bias will be supported if the animal shifts less by following inactivation of its associative sensory areas. 
 
Another promising avenue of research enabled by the perceptual gambling task is the study of confidence. Confidence is generally defined as the degree of belief in the truth of a proposition or the reliability of a piece of information, be it memory, observation or decision \citep{kepecsComputationalFrameworkStudy2012}. One important nuance is that there confidence has to be about a specific belief. In the PG task, one can distinguish the perceptual confidence from the decision confidence. 
In signal detection theory, perceptual confidence is defined as the distance from an internal representation of a stimulus to the decision boundary. On average, accuracy is a proxy for perceptual confidence  \citep{clarkeTwoTypesROC1959, galvinTypeTasksTheory2003}. This is a model of perceptual confidence, but it becomes indistinguishable from decision confidence in simple perceptual decision tasks. Recently, a Bayesian equivalent was proposed under the Bayesian decision theory, which defined confidence as the observer's posterior probability of being correct \citep{kepecsComputationalFrameworkStudy2012, hangyaMathematicalFrameworkStatistical2016, pougetConfidenceCertaintyDistinct2016a}. Using the Bayesian definition of confidence, research groups were able to find behavioral and neural correlates that resemble the qualitative signatures of Bayesian confidence in rats \citep{kepecsNeuralCorrelatesComputation2008a, lakOrbitofrontalCortexRequired2014} and humans \citep{sandersSignaturesStatisticalComputation2016}. We observe that the Bayesian decision theory framework can distinguish the two, such that the perceptual confidence is simply the log ratio of posterior \textit{sans} action cost, and the decision confidence is the log ratio of posterior with action cost. Our task makes investigating the neural underpinnings of these two kinds of confidence possible. Specifically, areas computing perceptual confidence should not vary across flashing conditions, as the perceptual boundary remains the same (unless, of course, the change in value causes a change in percept). In contrast, areas responsible for decision confidence should show differential activities across conditions, as the asymmetric reward would shift the indifference point, entailing a considerable change in decision confidence even with the same perceptual stimulus. Activities in these areas can be related to the Bayesian confidence measures as a test of theory, although it has been cautioned that the use of qualitative match is problematic \citep{adlerComparingBayesianNonBayesian2018}. 
 
In conclusion, we present the perceptual gambling task as a proof of concept, demonstrating that an integration of perceptual and value cues on a trial-by-trial basis is possible for rats. Future researchers interested in percept-value integration and confidence are encouraged to adopt this framework. Using model-based analysis, the brain circuits underlying these behavior can be rigorously explored with testable hypotheses. 
 
\section*{Acknowledgments}
We thank Yidi Chen, Cequn Wang, Anyu Fang, Yingkun Li and NengNeng Gao for technical assistance related to building and maintaining lab infrastructure as well as training animals.
 
\clearpage


\setcounter{figure}{0}
\renewcommand{\thefigure}{S\arabic{figure}}
\setcounter{table}{0}
\renewcommand{\thetable}{S\arabic{table}}




\begin{figure*}[htbp]
\includegraphics[width=0.9\linewidth]{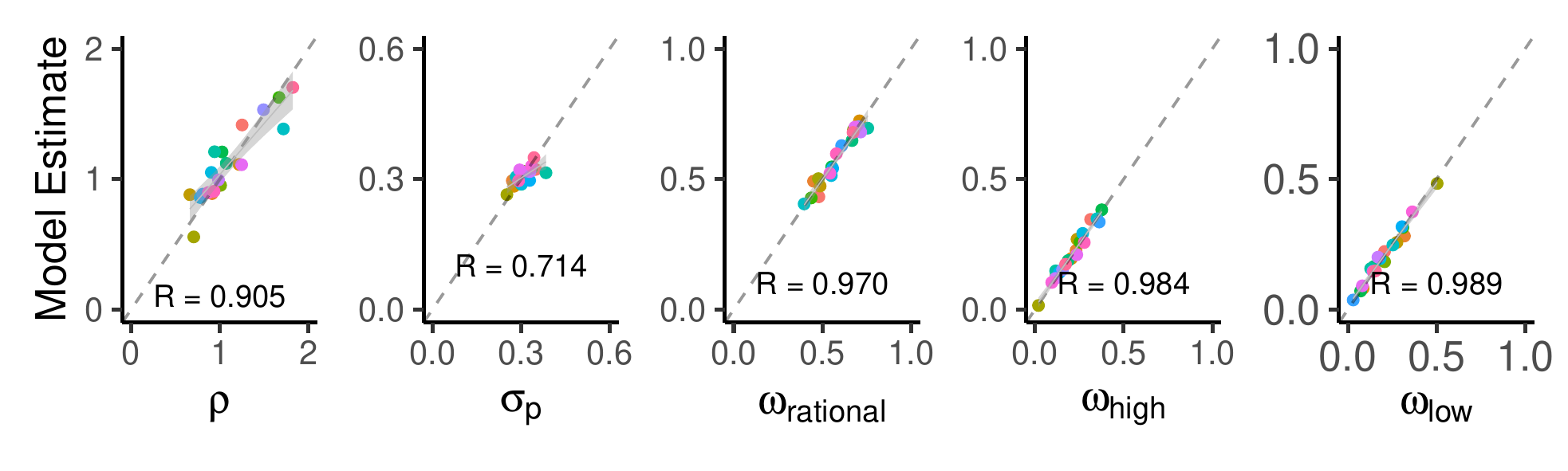}
\caption{The mixture BDT model can recover data-generating parameters accurately. Twenty Synthetic datasets were created by sampling from the same prior distributions as specified in Methods. The true parameter value is on the x-axis, the maximum \textit{a posteriori} estimation is on the y-axis. Color represents the identity of each synthetic dataset. All the parameters fall along the diagonal line (all $R > 0.7, p < 0.001$, Pearson's correlation test). 
}
\label{supp:sanity}
\end{figure*}

\begin{figure*}[htbp]
\includegraphics[width=0.9\linewidth]{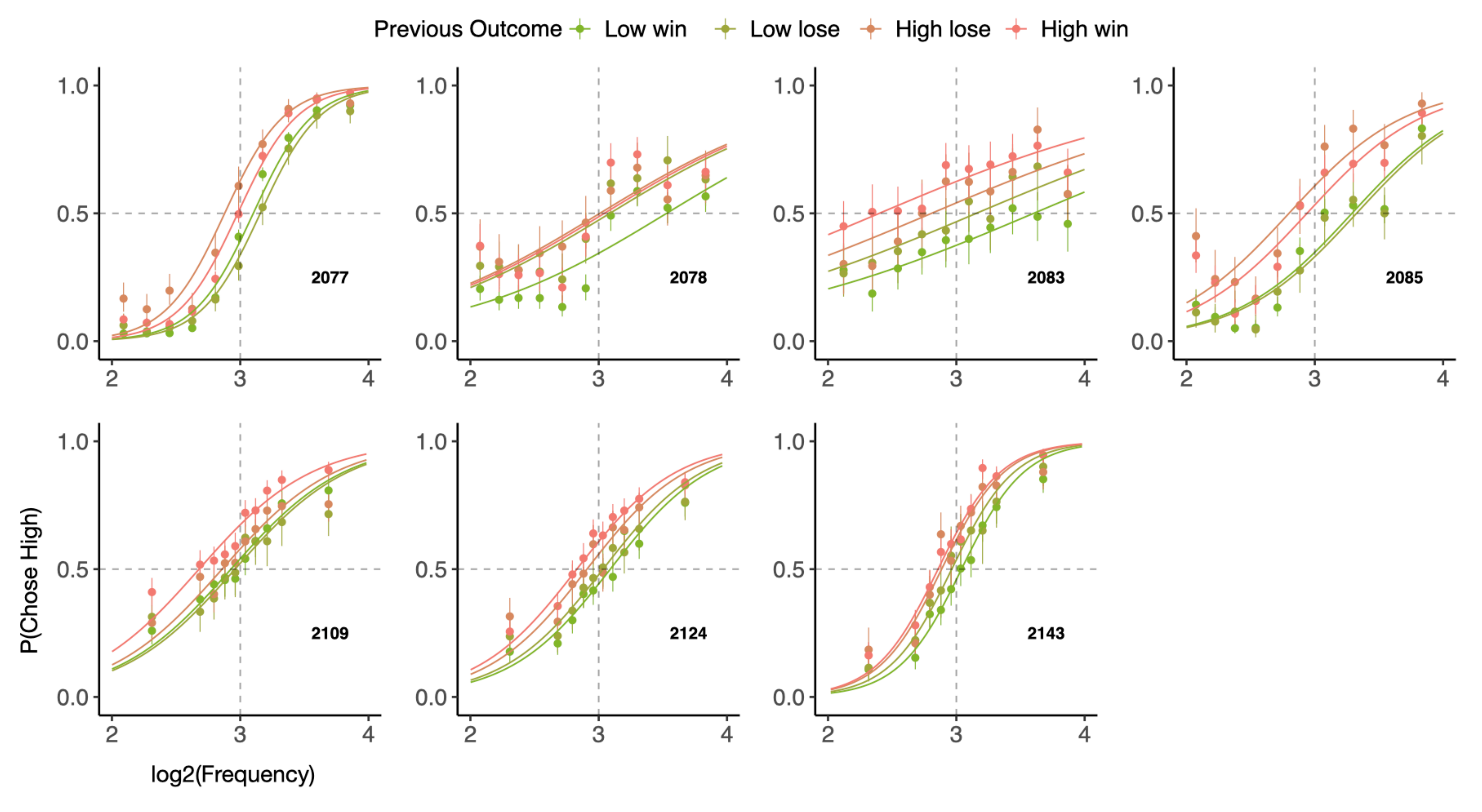}
\caption{Animal's choices were influenced by the outcome from the previous trial. The circles with error bars are the binned mean and 95\% binomial confidence intervals. The lines are generated by a generalized linear model. The colors represent the outcome of the previous trial.}
\label{supp:history}
\end{figure*}

\begin{figure*}[htbp]
\includegraphics[width=0.9\linewidth]{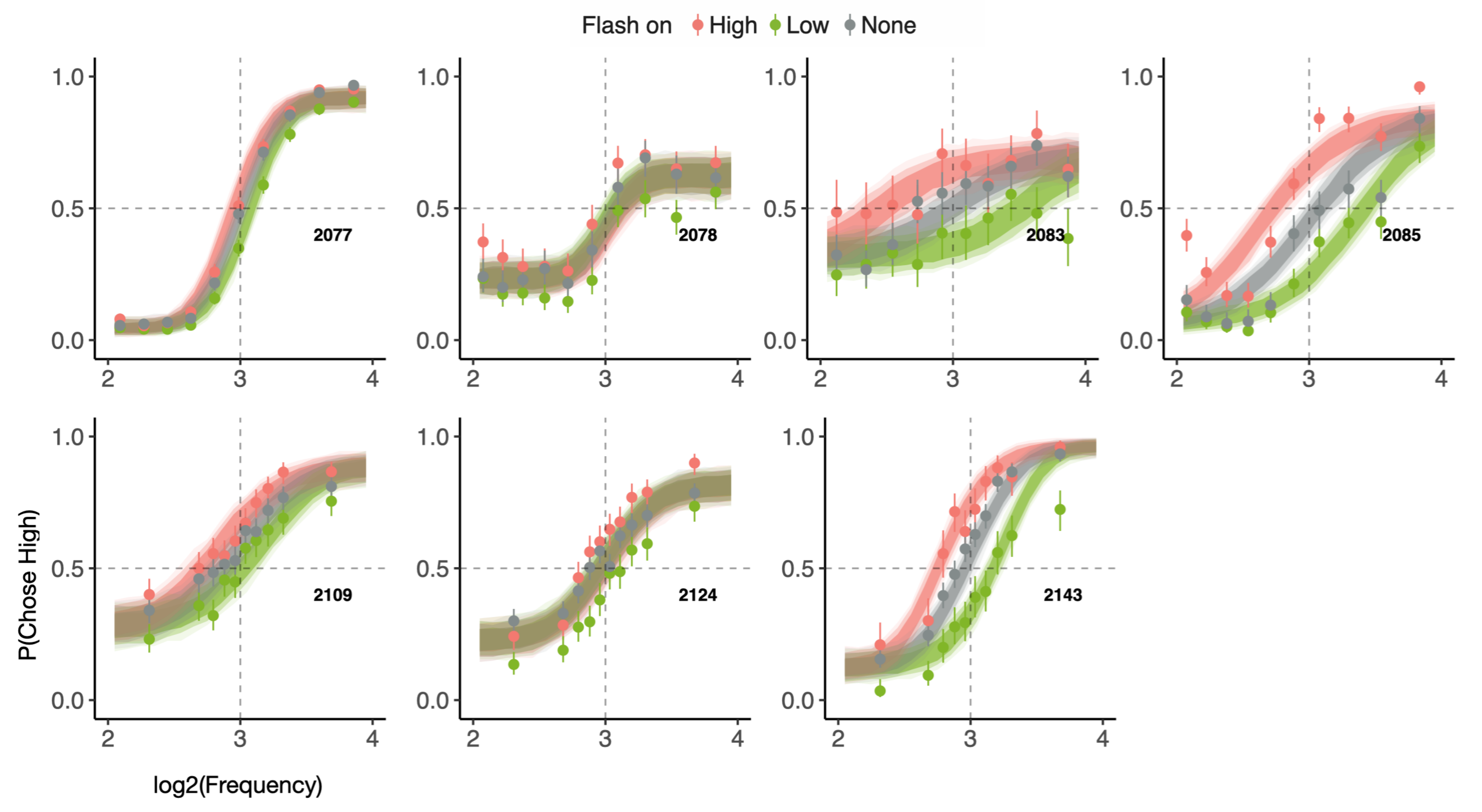}
\caption{The mixture-BDT model fits for each subject. The circles with error bars are the binned mean and 95\% binomial confidence intervals. The lines are generated from a synthetic sigmoid agent using maximum \textit{a posteriori} parameter estimates. The colors represent the flashing condition, with red = high side flash, green = low side flash, gray = no flash.
}
\label{supp:bdt_fits}
\end{figure*}

\begin{figure*}[htbp]
\includegraphics[width=0.9\linewidth]{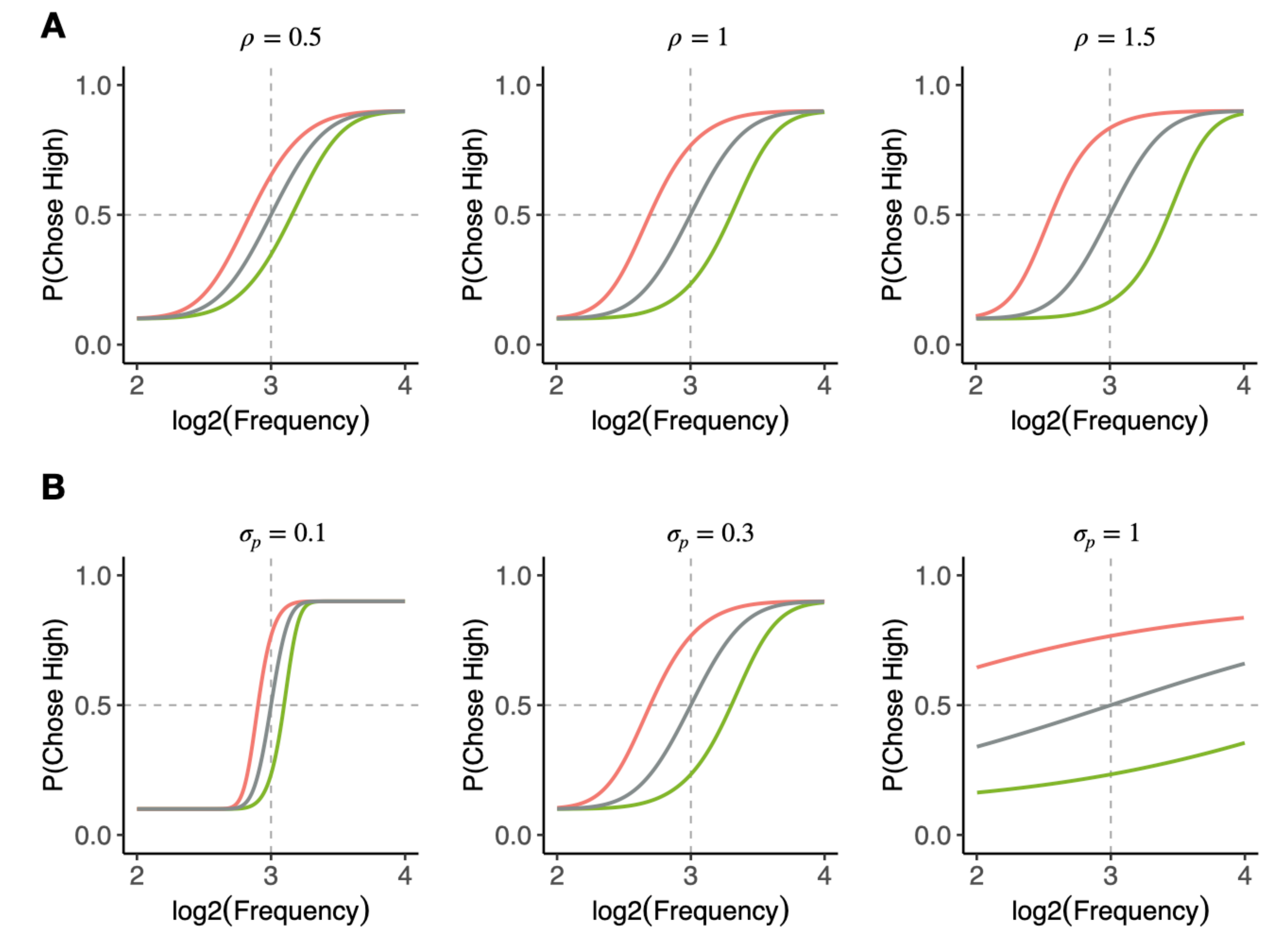}
\caption{The mixture-BDT model only allows for symmetric and horizontal shifts as a result of asymmetric reward. The data is simulated from a BDT agent with $\rho = 1$, $\sigma_p = 0.3$, $\omega_{rational} = 0.8$, $\omega_{high} = 0.1$, $\omega_{low} = 0.1$ unless otherwise specified. The colors represent the flashing condition, with red = high side flash, green = low side flash, gray = no flash.
\textbf{A.} Different levels of $\rho$, the utility exponent, determine the amount of horizontal shift without changing the slope. 
\textbf{B.} Different levels of $\sigma_p$, the perceptual noise, change both the amount of horizontal shift and slope of the psychometric functions. 
}
\label{supp:bdt_shift}
\end{figure*}

\begin{figure*}[htbp]
\includegraphics[width=0.9\linewidth]{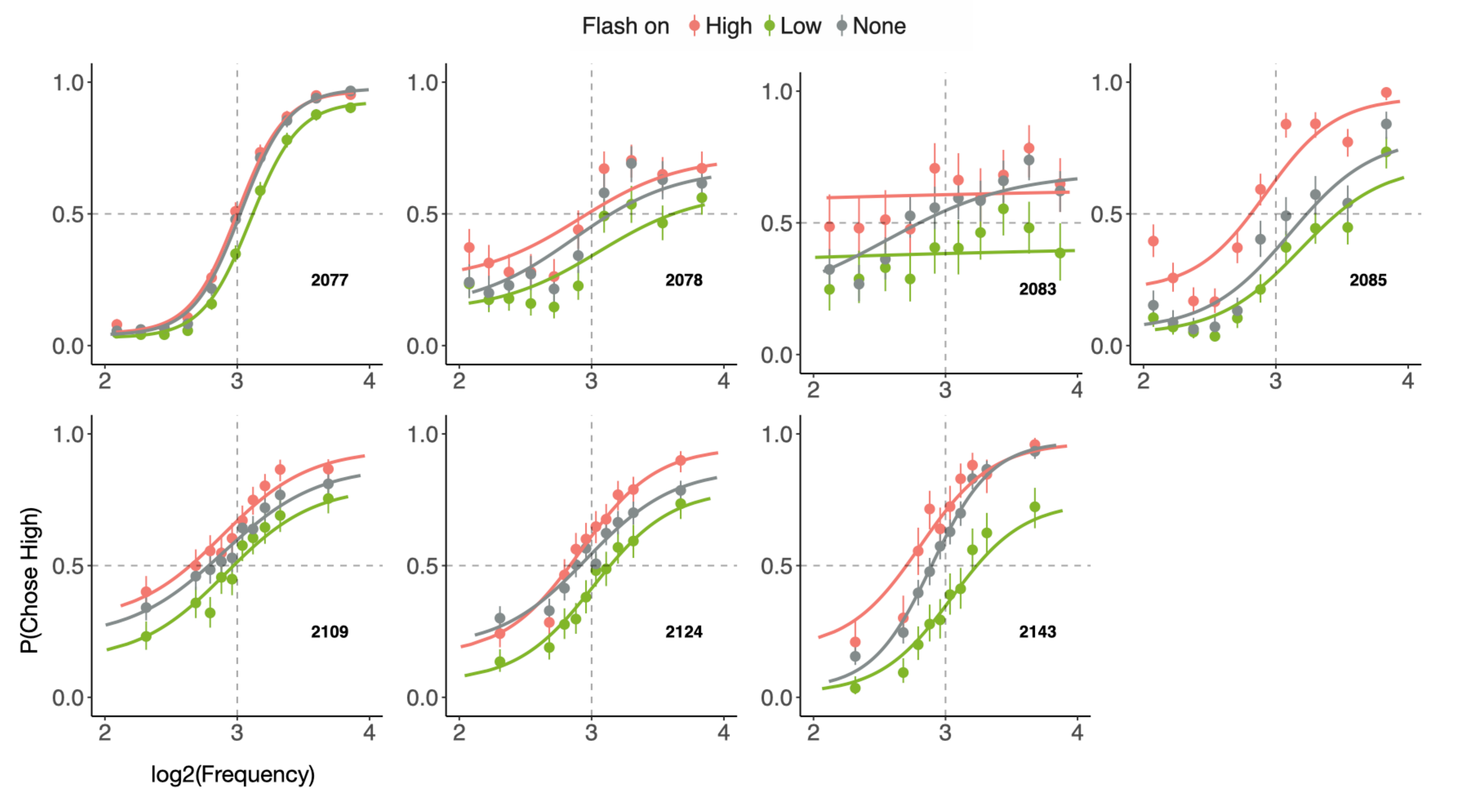}
\caption{Fits from the four-parameter sigmoid model for each subject. The circles with error bars are the binned mean and 95\% binomial confidence intervals. The lines are generated from a synthetic sigmoid agent using maximum \textit{a posteriori} parameter estimates. The colors represent the flashing condition, with red = high side flash, green = low side flash, gray = no flash.
}
\label{supp:sigmoid}
\end{figure*}

\clearpage

\bibliography{pg_2021}

\end{document}